\pdfoutput=1
\documentclass[12pt]{article}
\usepackage{amssymb,amsmath,mathtools,mathrsfs,slashed}
\usepackage{color,graphicx}
\usepackage{cite}
\usepackage{amssymb,amsmath}
\usepackage{multirow}
\usepackage{array}
\usepackage{hyperref}
\usepackage{comment}
\usepackage[export]{adjustbox}
\usepackage[top=3cm, bottom=3cm, left=2cm, right=2cm]{geometry}	

\usepackage{tabularx}
\newcolumntype{Y}{>{\centering\arraybackslash}X}
\usepackage{booktabs}
\AtBeginDocument{
    \heavyrulewidth=.08em
        \lightrulewidth=.05em
        \cmidrulewidth=.03em
        \belowrulesep=.65ex
        \belowbottomsep=0pt
        \aboverulesep=.4ex
        \abovetopsep=0pt
        \cmidrulesep=\doublerulesep
        \cmidrulekern=.5em
        \defaultaddspace=.5em
}
\usepackage{xcolor}

\newcommand\fverb{\setbox\pippobox=\hbox\bgroup\verb}
\newcommand\fverbdo{\egroup\medskip\noindent%
                              \fbox{\unhbox\pippobox}\ }
\newcommand\fverbit{\egroup\item[\fbox{\unhbox\pippobox}]}
\newbox\pippobox

\newcommand{\beq} {\begin{equation}}
\newcommand{\eeq} {\end{equation}}
\newcommand{\beqa} {\begin{eqnarray}}
\newcommand{\eeqa} {\end{eqnarray}}

\newcommand{\T}{\mathcal{T}}
\newcommand{\R}{\mathcal{R}}

\newcommand{\be}{\begin{equation}}
\newcommand{\ee}{\end{equation}}
\newcommand{\bea}{\begin{eqnarray}}
\newcommand{\eea}{\end{eqnarray}}

\newcommand{\vev}[1]{\left\langle#1\right\rangle}
\def\cN{{\cal N}}

\begin{document}
 
\begin{flushright}
HIP-2019-33/TH
\end{flushright}

\begin{center}

\centering{\Large {\bf Brane nucleation instabilities in non-AdS/non-CFT}}

\vspace{8mm}

\renewcommand\thefootnote{\mbox{$\fnsymbol{footnote}$}}
Oscar Henriksson,${}^{1,2}$\footnote{oscar.henriksson@helsinki.fi}
Carlos Hoyos,${}^{3,4}$\footnote{hoyoscarlos@uniovi.es} and
Niko Jokela${}^{1,2}$\footnote{niko.jokela@helsinki.fi}

\vspace{4mm}
${}^1${\small \sl Department of Physics} and ${}^2${\small \sl Helsinki Institute of Physics} \\
{\small \sl P.O.Box 64} \\
{\small \sl FIN-00014 University of Helsinki, Finland}

\vspace{2mm}
\vskip 0.2cm
${}^3${\small \sl Department of Physics} \\
{\small \sl Universidad de Oviedo} \\
{\small \sl c/ Federico Garc\'{\i}a Lorca 18, ES-33007 Oviedo, Spain} 

\vspace{2mm}
\vskip 0.2cm
${}^4${\small \sl Instituto Universitario de Ciencias y Tecnolog\'{\i}as Espaciales de Asturias (ICTEA)}\\
{\small \sl  Calle de la Independencia, 13, 33004 Oviedo, Spain}

\end{center}

\vspace{0mm}

\setcounter{footnote}{0}
\renewcommand\thefootnote{\mbox{\arabic{footnote}}}

\begin{abstract}
\noindent 
We speculate that the weak gravity conjecture applied to theories with holographic duals bans the existence of disordered phases at zero temperature. We test this idea by introducing a non-zero baryon chemical potential in a deformation of the $SU(N_c)\times SU(N_c)$ Klebanov-Witten gauge theory with broken supersymmetry and conformal invariance. At low temperature, a disordered phase dual to a black brane geometry is unstable for low chemical potentials and metastable for high values. In the metastable phase, states with a partial Higgsing of the gauge group are favored over the normal disordered phase. This is reflected in the properties of the effective potential for color branes in the dual geometry, where the appearance of a global minimum outside the horizon signals the onset of a brane nucleation instability. When the Higgsing involves only one of the group factors, the global minimum remains at a finite distance from the horizon, making it possible to construct holographic duals to metastable ``color superconducting'' states. We also consider branes dual to excitations with baryon charge, but find that the extremal geometry remains marginally stable against the emission of particles carrying baryon charge independently of the strength of the deformation.
\end{abstract}

\newpage

\tableofcontents


\newpage


\section{Introduction}\label{sec:introduction}

Hot matter is typically in a very uniform and symmetric -disordered- phase, a plasma or other type of fluid that can be effectively described using hydrodynamics.  Cold matter, on the other hand, can manifest an endless variety of forms with different types of order, as the richness of states studied in condensed matter physics show \cite{Wen:2019qtq}. The same trend applies to more fundamental theories, such as QCD. At high temperature, the quark-gluon matter as observed in heavy ion collisions is reminiscent of a plasma phase (see \cite{Brambilla:2014jmp} for a review). At low temperature, and at ultra-high densities, matter as described by perturbative QCD is believed to organize into the color-flavor locking (CFL) phase \cite{Alford:1998mk} (see also \cite{Alford:2007xm,Brambilla:2014jmp} for reviews). Contrarily, at intermediate densities, the difference between the quark masses starts to be relevant and there are several possible phases that could be realized, including phases that break spacetime symmetries. Examples include an anisotropic phase consisting of a Kaon condensate with spontaneously generated currents (known as currCFL-K0) \cite{Schafer:2005ym,Kryjevski:2008zz} and phases showing spontaneous breaking of translation invariance, forming a crystalline CFL \cite{Alford:2000ze,Rajagopal:2006ig,Mannarelli:2006fy}. It is also important to recall that even the large-$N_c$ limit of high density QCD is assumed to be in a symmetry-broken phase, albeit different from CFL, the chiral density wave (CDW)\cite{Deryagin:1992rw,Shuster:1999tn}.

It is nowadays apparent that strongly coupled theories with gravity duals also evolve from disordered to ordered phases as they are cooled down and exhibit a similar richness. At high temperature, the ground state is typically in a uniform plasma phase whose gravity dual is a black hole geometry and whose fluctuations are effectively captured by relativistic hydrodynamics \cite{Policastro:2002se,Policastro:2002tn,Bhattacharyya:2008jc}. As the temperature is lowered, the disordered phase can become unstable towards the breaking of symmetries, either  internal \cite{Hartnoll:2008vx}, spacetime \cite{Domokos:2007kt,Nakamura:2009tf}, or both \cite{Aharony:2007uu,Gubser:2008wv,Ammon:2008fc}. There is a whole zoo of ordered phases that include superfluids  \cite{Hartnoll:2008vx,Hartnoll:2008kx,Horowitz:2009ij,Aharony:2007uu,Gubser:2008wv,Ammon:2008fc}, anisotropic states \cite{Azeyanagi:2009pr,Mateos:2011ix,Mateos:2011tv,Giataganas:2012zy,Rougemont:2014efa,Fuini:2015hba,Conde:2016hbg,Gursoy:2016ofp,Penin:2017lqt,Bea:2017iqt,Giataganas:2017koz,Itsios:2018hff,Jokela:2019tsb,Gran:2019djz}, striped phases \cite{Ooguri:2010kt, Ooguri:2010xs, Bayona:2011ab, Bergman:2011rf, Jokela:2012vn, BallonBayona:2012wx, Bu:2012mq, Jokela:2012se, Rozali:2012es, Withers:2013loa, Withers:2013kva, Rozali:2013ama, Ling:2014saa, Jokela:2014dba, Amoretti:2016bxs,Cremonini:2017usb,Jokela:2017fwa,Jokela:2016xuy,Cremonini:2017qwq,Jokela:2017ltu,Amoretti:2017frz,Amoretti:2017axe,Donos:2018kkm,Gouteraux:2018wfe,Li:2018vrz}, and even color superconducting phases  \cite{Chen:2009kx,Basu:2011yg,Rozali:2012ry,BitaghsirFadafan:2018iqr,Faedo:2018fjw,Henriksson:2019zph}.

The instability of the disordered phase is sometimes subtle, being present in the full string theory even in cases where the pure gravity solution may look stable. A prime example is brane nucleation, by which a bound state of branes becomes unstable and starts to shed some of its components, as first described in \cite{Seiberg:1999xz} (see \cite{Kleban:2004bv} for a clear overview). In the context of gauge/gravity duality this kind of process has also been dubbed  as ``Fermi seasickness''  \cite{Hartnoll:2009ns,McInnes:2009zp} and was applied in an AdS/QCD approach to the phase diagram in some previous works  \cite{McInnes:2009zp,McInnes:2009ux}. An interesting string theory example of brane nucleation occurs in the charged black branes studied in \cite{Herzog:2009gd}. These geometries are dual to disordered states with finite baryon density in the Klebanov-Witten (KW) theory \cite{Klebanov:1998hh}, which is a $(3+1)$-dimensional CFT. Physics thus depends only on the ratio of the two relevant scales, temperature and chemical potential. There seems to be no obvious instabilities in the classical gravity solution, but ``color'' branes, with a worldvolume parallel to the horizon, feel an effective potential that allows them to escape from the horizon to infinity at low enough temperature. Interestingly, a similar mechanism has been shown to exist even in ${\cal{N}}=4$ super Yang-Mills at finite $R$-charge chemical potential \cite{Henriksson:2019zph}.

It shold be noted, however, that there are other cases in which even the brane nucleation instability is absent, such as the $(2+1)$-dimensional CFTs at non-zero charge of \cite{Klebanov:2010tj} or the $(3+1)$-dimensional theories with quenched flavors and finite baryon charge of \cite{Ammon:2012mu}. In both cases it is found that there is a classical moduli space in the extremal limit (although the quenched approximation in the second case is expected to break down \cite{Bigazzi:2013jqa}). As these solutions are not supersymmetric, quantum or stringy corrections may lift the moduli space and still render the classical symmetric solution unstable.  

Another interesting aspect of these kind of theories is that the gravity duals \cite{Herzog:2009gd,Klebanov:2010tj} contain charged particles in the form of wrapped (``baryonic'') branes whose mass to charge ratio becomes critical in the extremal limit (but it is above the critical value outside extremality). Assuming the mass to charge ratio of other states is larger or equal, this makes the extremal black branes marginally stable with respect to emission of charged particles. According to the weak gravity conjecture (WGC) of \cite{ArkaniHamed:2006dz}, quantum corrections should decrease the relative mass to charge ratio and render the extremal black branes unstable.\footnote{Some arguments have recently been given that the WGC holds in holographic models, see, {\emph{e.g.}}, \cite{Montero:2018fns}. The implications for non-supersymmetric vacua with AdS duals were discussed in  \cite{Ooguri:2016pdq}.} Assuming the WGC holds, this implies that {\em theories with a holographic dual at non-zero charge density have no disordered phases at zero temperature}, barring external sources of disorder. 

One of our goals is to further explore the phase diagram of the KW theory by introducing an explicit breaking of conformal invariance in the form of a mass term for the scalar components of the chiral multiplets. This introduces a new mass scale in the theory, and the phase diagram then depends on the ratios of both temperature and chemical potential to the new scale. This is of interest for several reasons, one of them is to check if brane nucleation is still the mechanism by which near-extremal black branes become unstable. Another reason is that models where finite baryon density can be introduced without introducing flavor branes (with or without backreaction) are rarely studied, but very interesting because they do not require additional approximations such as quenching or smearing. In particular, non-conformal theories may serve as a theoretical laboratory to study high density quark or nuclear matter, similar to the one expected to be found in the interior of neutron stars. In fact, recent work shows that holographic models can have phenomenologically viable equations of state \cite{Hoyos:2016cob,Ecker:2017fyh,Jokela:2018ers,Ishii:2019gta,Chesler:2019osn} and can be used to model stars that satisfy existing observational constraints \cite{Hoyos:2016zke,Annala:2017tqz,Jokela:2018ers,Ecker:2019xrw}, but much work still needs to be done. 

The paper is organized as follows. In Sec.~\ref{sec:setup} we review the KW field theory and its holographic dual, as well as the five-dimensional truncation we employ. In Sec.~\ref{sec:thermo} we construct the black brane geometries dual to the disordered phase and work out some of their thermodynamic properties. We then proceed to discuss brane nucleation instabilities, which can be found by computing the effective potential for probe branes in the background geometry. In Sec.~\ref{sec:color} we probe the geometry with an additional color D3-brane, as well as with a ``color'' D5-brane, in both cases finding an instability analogous to color superconductivity at low temperature. In Sec.~\ref{sec:baryon} we then consider the effective potential felt by baryonic D3-brane probes. Sec.~\ref{sec:discussion} briefly summarizes our thoughts. We supplement the paper with two appendices which contain computational details complementing the discussion in the bulk part of the text.


\section{Deformation of the Klebanov-Witten theory}\label{sec:setup}

The KW theory is a superconformal field theory (SCFT) that emerges as the low energy effective theory of D3-branes placed at the singularity of the conifold with base $T^{1,1}=(SU(2)\times SU(2))/U(1)$. The theory is $\cN=1$ super Yang-Mills with gauge group $SU(N_c)\times SU(N_c)$, where $N_c$ is the number of D3-branes. In addition to the gauge fields, there are two sets of chiral multiplets $A_\alpha$, $B_{\tilde\alpha}$, $\alpha,\tilde\alpha=1,2$, in the $(N_c,\bar{N}_c)$ and $(\bar{N}_c,N_c)$ bifundamental representations, respectively, each of them a doublet of a different global $SU(2)$ symmetry. The chiral multiplets have charge $1/2$ under a non-anomalous $U(1)_R$ symmetry, and there is an additional global $U(1)_B$ baryon symmetry
\be
A_\alpha \to e^{i\theta} A_\alpha \ , \ B_{\tilde\alpha}\to e^{-i\theta} B_{\tilde\alpha} \ .
\ee
The exactly marginal superpotential that preserves these symmetries is
\be
 W=\frac{\lambda}{2}\epsilon^{\alpha\beta}\epsilon^{\tilde\alpha\tilde\beta} {\rm Tr}\,\left( A_\alpha B_{\tilde\alpha} A_\beta B_{\tilde\beta}\right) \ .
\ee
Note that there is also a discrete $\mathbb{Z}_2$ symmetry that simultaneously exchanges the two gauge groups and the $A$ and $B$ multiplets.

Conformal invariance can be broken by adding additional terms to the potential. This can be done in a controlled way by taking the scalar component of a BPS operator, whose conformal dimension is protected. In our case we take the deformation to be quadratic in the scalar components $a$ and $b$ of the chiral multiplets $A$ and $B$:
\be\label{eq:massAB}
V_M=\pm M^2 {\rm Tr}\,\left(a^\dagger \cdot a-b^\dagger\cdot  b \right) \ .
\ee
Here $M$ is a parameter with dimension of mass. This term preserves $SU(2)\times SU(2)$ invariance as well as $U(1)_R$ and $U(1)_B$ symmetries. It breaks the discrete $\mathbb{Z}_2$ symmetry and, as we are not including analogous terms for the fermion components, supersymmetry is also broken. Note that for either sign the moduli space of the theory is lifted and in fact the classical potential is unbounded from below. Therefore, the theory with this deformation does not have a well-defined ground state. Nevertheless, in principle it is possible to make sense of this theory if we turn on a finite temperature $T$. In that case we expect that the scalar fields acquire effective masses and the effective potential at quadratic order becomes
\be\label{eq:VmT}
V_{M,T}\simeq (c_A T^2\pm M^2) {\rm Tr}\,\left(a^\dagger \cdot a\right)+(c_B T^2\mp M^2) {\rm Tr}\,\left(b^\dagger \cdot b\right) \ ,
\ee
where $c_A,c_B>0$ can be determined at weak coupling by a one-loop calculation. The effective potential in this case is bounded from below as long as the temperature is large enough $c_A T^2>M^2$ or $c_B T^2 >M^2$.

In addition to this deformation we will consider states with non-zero baryon and $R$-charge. Let us define the ``current'' operators
\be\label{eq:jajb}
J_a^\mu =\frac{i}{2} {\rm Tr}\, \left(a^\dagger \cdot \overset{\leftrightarrow}{D^\mu} a\right),\ \ J_b^\mu =\frac{i}{2} {\rm Tr}\,\left(b^\dagger \cdot \overset{\leftrightarrow}{D^\mu} b\right) \ ,
\ee
where $D_\mu$ are the appropriate covariant derivatives acting on the scalars.  The baryon and $R$-charge currents are
 \be\label{eq:currents}
 J_B^\mu=J_a^\mu-J_b^\mu+\text{fermions} \ , \ J_R^\mu=\frac{1}{2}\left( J_a^\mu+J_b^\mu\right)+\text{fermions} \ .
 \ee
The chemical potentials for the baryon ($\mu_B$) and  $R$-charges ($\mu_R$) are naturally incorporated by adding to the potential a term of the form
\be
V_ \mu=-\mu_R J_R^0-\mu_B J_B^0 \ .
\ee
In this expression the covariant derivative appearing in \eqref{eq:jajb} must include a coupling to the chemical potentials, that enter similarly to background fields
\be
D_0 a\to \left(D_0-i\mu_B-\frac{i}{2}\mu_R\right)a\ , \ D_0 b\to \left(D_0+i\mu_B-\frac{i}{2}\mu_R\right)b \ .
\ee
Taking into account the kinetic term, the terms linear in derivatives can be removed by factoring out a phase from the scalar fields
\be
a=e^{i\left(\mu_B+\frac{\mu_R}{2}\right) t} \tilde{a}\ , \ b=e^{i\left(-\mu_B+\frac{\mu_R}{2}\right) t} \tilde{b} \ .
\ee
In this case
\be
V_ \mu=-\left( \mu_B+\frac{\mu_R}{2}\right)^2{\rm Tr}\left(\tilde{a}^ \dagger\cdot \tilde{a}\right)-\left( \mu_B-\frac{\mu_R}{2}\right)^2{\rm Tr}\left(\tilde{b}^ \dagger\cdot \tilde{b}\right) \ .
\ee
Combined with \eqref{eq:VmT}, the total effective potential at quadratic order is
\be
V_{M,T,\mu}\simeq \left[c_A T^2\pm M^2-\left( \mu_B+\frac{\mu_R}{2}\right)^2\right] {\rm Tr}\,\left(\tilde{a}^\dagger \cdot \tilde{a}\right)+\left[c_B T^2\mp M^2-\left( \mu_B-\frac{\mu_R}{2}\right)^2\right] {\rm Tr}\,\left(\tilde{b}^\dagger \cdot \tilde{b}\right) \ .
\ee
For large enough chemical potentials the quadratic potential is unbounded from below even if $M=0$; a similar instability has been discussed at length for the $R$-charge chemical potentials of $\mathcal{N}=4$ SYM \cite{Yamada:2006rx,Yamada:2007gb,Yamada:2008em,Hollowood:2008gp,Henriksson:2019zph}. In that case the large chemical potential instability manifested itself in the holographic dual as a brane nucleation instability. In the case at hand we expect that a similar identification can be done. However, the $M\neq 0$ instability could either be related to brane nucleation or to an instability of the dual geometry at the level of classical gravity. We will discuss these points in more detail later on.

\subsection{Holographic dual}

In the large-$N_c$ limit, the KW theory has a dual description in terms of a weakly coupled type IIB string theory on a manifold which is a direct product of an asymptotically AdS$_5$ (aAdS$_5$) spacetime and $T^{1,1}$. At strong 't Hooft coupling, classical type IIB supergravity provides the leading order approximation to the properties of the theory. The  $M\neq 0$ and baryon charge sector is captured by a consistent supersymmetric truncation to five dimensions \cite{Cassani:2010na}, which greatly simplifies the problem of finding the dual geometries to the deformed KW theory. We will focus on disordered states at non-zero temperature and charge, with holographic duals that are charged black brane geometries. Similar geometries were constructed in \cite{Herzog:2009gd} at zero mass $M=0$, using a subset of the supersymmetric truncation. Along the way, comparison with their results will be used as a check of our analysis.

The details of the truncation is in Appendix~\ref{app:sugra}. The action for the five-dimensional truncated theory is
\begin{equation}\label{eq:5DLag}
 S_{5D} = \int d^5 x \sqrt{-g} \mathcal{L}_{5D} + S_{CS} \, ,
\end{equation}
where
\begin{align}
 \begin{split}
 \mathcal{L}_{5D} =& R - \frac{10}{3}(\partial_{\mu}\chi)^2 - 5(\partial_{\mu}\eta)^2 - (\partial_{\mu}\lambda)^2 - V \\
 &-\frac{1}{4}e^{2\eta-\frac{4}{3}\chi}\left[ \cosh(2\lambda) \left( (F_{\mu\nu})^2 + (F^M_{\mu\nu}-F^R_{\mu\nu})^2 \right) - 2\sinh(2\lambda)(F^M_{\mu\nu}-F^R_{\mu\nu})F^{\mu\nu} \right] \\
 &-\frac{1}{8}e^{-4\eta+\frac{8}{3}\chi}(F^R_{\mu\nu})^2 - 4e^{-4\eta-4\chi}(A^M_{\mu})^2 \, ,
 \end{split}
\end{align}
and the potential is
\begin{equation}
 V = 8e^{-\frac{20}{3}\chi} + 4e^{-\frac{8}{3}\chi} (e^{-6\eta}\cosh(2\lambda)-6e^{-\eta}\cosh(\lambda)) \, ;
\end{equation}
we have set the radius of curvature $L=1$. The Chern-Simons term is
\begin{equation}
 S_{CS} = \frac{1}{2\sqrt{2}} \int (A_M - A_R) \wedge F_M \wedge F_R - \frac{1}{2\sqrt{2}} \int A \wedge F \wedge F_R \, .
\end{equation}
The potential has a critical point at $\lambda=\chi=\eta=0$. If the gauge fields are also set to zero, the solution to the equations of motion is an AdS$_5$ geometry (of radius $L=1$), dual to the KW theory at the origin of the moduli space where the theory enjoys the full conformal invariance. Expanding around this point
\be
\lambda=\frac{1}{\sqrt{2}} \delta\lambda \ , \ \chi=\sqrt{\frac{3}{20}} \delta\chi \ , \ \eta=\frac{1}{\sqrt{10}} \delta \eta \ ,
\ee
the action for the scalars to quadratic order is
\be
\mathcal{L}_S\simeq -\frac{1}{2}\left[ (\partial_\mu\delta\lambda)^2+(\partial_\mu\delta\chi)^2+(\partial_\mu\delta\eta)^2-4\delta\lambda^2+32 \delta\chi^2+12\delta\eta^2\right] \ .
\ee
Therefore, the scalar fields around the critical point have masses $m^2_\lambda L^2=-4$, $m^2_\chi L^2=32$, $m^2_\eta L^2=12$. Following the usual AdS/CFT dictionary, we can identify $\lambda$ as the field dual to the operator of conformal dimension $\Delta=2$  \eqref{eq:massAB},\footnote{This identification is also based on symmetries, $\lambda$ belongs to a $SU(2)\times SU(2)$ invariant truncation and is not charged under the field $a_\mu^R$ dual to $U(1)_R$ current. The only other candidate ${\rm Tr}\,(|a|^2+|b|^2)$ is not a BPS operator.} while $\chi$ and $\eta$ are dual to scalar operators of dimensions $\Delta=8$ and $\Delta=6$, respectively.

In order to identify the operators dual to the vector fields we should also expand the action to quadratic order around the critical point. Note that as the kinetic terms are mixed, we will diagonalize the quadratic action defining
\be
A^R_\mu=\frac{2}{\sqrt{3}}\left( \frac{1}{\sqrt{2}}a^R_\mu+a^M_\mu\right),\ \ A^M_\mu=\sqrt{3} a^M_\mu \ .
\ee
The action for the vector fields becomes
\be
\mathcal{L}_V\simeq -\frac{1}{4} (F_{\mu\nu})^2-\frac{1}{4}(\partial_\mu a^R_\nu-\partial_\nu a^R_\mu)^2-\frac{1}{4}(\partial_\mu a^M_\nu-\partial_\nu a^M_\mu)^2-12 (a^M_\mu)^2 \ .
\ee
The massless vector fields $A_\mu$ and $a_\mu^R$ are dual to the baryon and $R$ currents in \eqref{eq:currents}. The vector field $a^M_\mu$ has mass $m_M^2L^2=24$, so it is dual to a vector operator of conformal dimension $\Delta=7$. 

We will allow for configurations that flow in the UV to a fixed point, this means that we will bar sources for the irrelevant operators dual to the scalars $\chi$ and $\eta$, and for the massive vector field $a^M_\mu$. In the holographic dual the metric will approach $AdS_5$ close to the asymptotic boundary $r\to \infty$:
\be
ds_{5}^2 \simeq \frac{L^2}{r^2}dr^2+\frac{r^2}{L^2}\eta_{\mu\nu}dx^\mu dx^\nu \ ,
\ee
while the fields dual to irrelevant operators vanish
\be
\chi\sim \frac{1}{r^8} \ , \ \eta\sim \frac{1}{r^6},\ \ a_\mu^M\sim \frac{1}{r^6} \ .
\ee
An explicit breaking of conformal invariance will be realized by introducing a coupling $\sim M^2$ to the $\Delta=2$ operator. On the gravity side, the dual scalar field will have an asymptotic expansion of the form
\be
\lambda \sim \frac{L^4M^2}{r^2}\log\frac{r}{L} \ .
\ee
Non-zero baryonic and $R$-charge chemical potentials can be introduced by turning on the time components of the massless vector fields. They can be defined as the integral of the radial electric flux between the black brane horizon of the dual geometry and the asymptotic boundary 
\be
\mu_B=\int_{r_H}^\infty dr F_{rt}\ , \ \mu_R=\int_{r_H}^\infty dr (\partial_ra_t^R-\partial_t a_r^R) \ .
\ee
We will work with stationary solutions, and thermal equilibrium usually demands that the vector fields vanish at the horizon, so that the chemical potentials coincide with the values of the vector fields at the boundary
\be
\mu_B=\lim_{r\to\infty} A_t\ , \ \mu_R=\lim_{r\to \infty} a_t^R \ .
\ee 
With this we have all the necessary ingredients to construct solutions to the five-dimensional action that are dual to finite temperature and charge density states, with the conformal symmetry breaking coupling $M$ turned on.


\section{Black brane geometries and thermodynamics}\label{sec:thermo}

The five-dimensional action \eqref{eq:5DLag}  admits a family of black brane solutions based on the following Ansatz for the metric and vector fields
\begin{equation}
\begin{split}
&ds_5^2 = -g e^{-w} dt^2 + \frac {dr^2}{g} + \frac {r^2} {L^2} \sum_{i=1}^3 (d x^i)^2 \\
& A=\Phi(r)dt \ , \ A_R=\Phi_R(r)dt \ , \  A_M=\Phi_M(r)dt \ .
\end{split}
\end{equation}
The scalar fields are also non-trivial, depending on the radial coordinate: $\lambda(r)$, $\chi(r)$, $\eta(r)$. Inserting this Ansatz into the equations of motion derived from the action \eqref{eq:5DLag} gives a system of eight differential equations involving eight functions of $r$. This system is first order in the derivatives of the metric functions $g(r)$ and $w(r)$ and second order in the other functions. It is fairly complicated, but we can simplify it by noting that the equations for the two massless gauge fields can be integrated, allowing us to replace them with two first order equations written in terms of two integration constants $\mathcal{Q}_B$ and $\mathcal{Q}_R$:
\begin{equation}\label{eq:QBQR}
\begin{split}
 \mathcal{Q}_B =& \frac{e^{\frac{w(r)}{2}+2\eta(r)-\frac{4\chi(r)}{3}}r^3}{16\pi G_5}\left[\cosh(2\lambda(r))\Phi'(r) + \sinh(2\lambda(r))\left(\Phi_R'(r) - \Phi_M'(r)\right)\right] \\
 \mathcal{Q}_R =& \frac{e^{\frac{w(r)}{2}-4\eta(r)-\frac{4\chi(r)}{3}}r^3}{32\pi G_5}\left[2e^{6\eta(r)}\left( \cosh(2\lambda)\left(\Phi_R'(r) - \Phi_M'(r)\right) + \sinh(2\lambda)\Phi'(r)\right) + e^{4\chi(r)}\Phi_R'(r) \right] \ . 
 \end{split}
\end{equation}
The constants $\mathcal{Q}_B$ and $\mathcal{Q}_R$ will be related to the baryonic and $R$-charge densities below. 

We thus end up with a simpler system of four first order and four second order differential equations that we wish to solve numerically. In order to do this, we will employ a double-sided shooting method: We expand the equations both near the aAdS boundary at $r\rightarrow\infty$ and near the black brane horizon at $r=r_H$. Each of these expansions leaves us with some free parameters. By choosing initial values for these, we can numerically solve the system of equations by integrating from both the boundary and the horizon up to some midpoint in the bulk, say $r=r_0$. At this midpoint we compute the vector
\begin{equation}\label{eq:midpointVector}
 X \equiv \left\{ g, w, \Phi, \Phi_R, \Phi_M, \Phi_M', \eta, \eta', \chi, \chi', \lambda, \lambda' \right\}
\end{equation}
for both solutions. We then require that the difference between the two $X$'s so computed is zero, in order to have a well-behaved solution in the whole region between the boundary to the horizon:
\begin{equation}\label{eq:matching}
 \Big|X(r_0)|_{horizon\rightarrow bulk} - X(r_0)|_{boundary\rightarrow bulk}\Big| < \rm{small} \ .
\end{equation}
This last step is accomplished by using the \textbf{FindRoot} function in \textsl{Mathematica} to tune the free parameters until \eqref{eq:matching} is satisfied. Note that in \eqref{eq:midpointVector} we only needed to include the first derivatives of the functions whose \emph{second} derivatives appear in the system of equations.

To check that \eqref{eq:matching} fixes all parameters, we briefly discuss the two expansions. For $r\to \infty$ we need to impose that the spacetime really is asymptotically AdS$_5$ --- this means the different functions should have the asymptotic behavior given in \eqref{eq:UVexpansion}. This leaves us with ten unfixed parameters in the near-boundary expansion. On the other hand, near the black brane horizon, which we fix to be at $r=r_H=1$, time components of the metric and the vector fields should vanish in order to have a smooth continuation to Euclidean signature, as expected for a geometry dual to a state at thermal equilibrium:
\be
g(r)\sim \Phi(r)\sim \Phi_R(r)\sim \Phi_M(r) \sim O(r-r_H) \ .
\ee
The other functions should be regular at the horizon. Imposing this, we are left with seven unfixed parameters in the near-horizon expansion of the equations of motion. There is some overlap in the expansions, since $\mathcal{Q}_B$ and $\mathcal{Q}_R$ are part of the parameters in both cases. Thus, we have a total of fifteen independent parameters from these expansions. As will be discussed more below, we elect to be in a \emph{mixed ensemble}: grand canonical for the baryon symmetry, fixing the chemical potential $\mu_B$ corresponding to $\Phi_{0,0}$ in \eqref{eq:UVexpansion}, and canonical for the $R$-symmetry, fixing the charge density which is given by $\mathcal{Q}_R$. Lastly, we fix the source of the scalar dual to $\lambda$, corresponding to $\lambda_{2,1}$ in \eqref{eq:UVexpansion}. This takes us down to twelve parameters, which matches exactly with the number of conditions imposed by \eqref{eq:matching}.

\begin{figure}
\begin{center}
\includegraphics[scale=0.62]{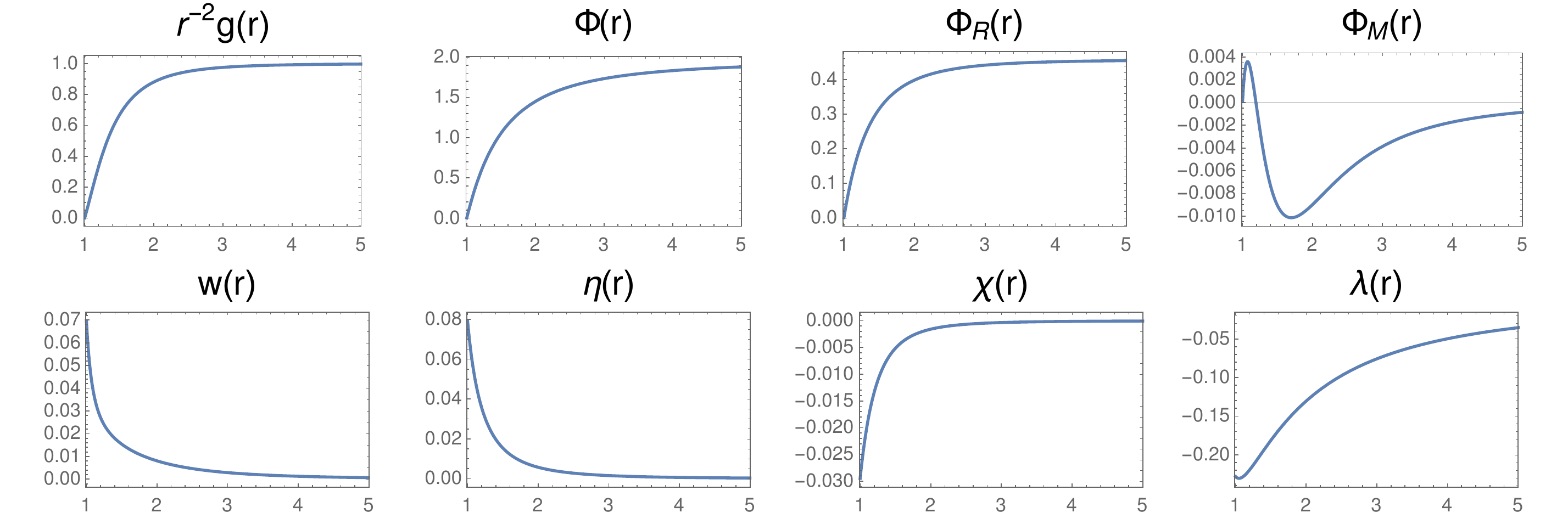}
\caption{An example solution, at $T/M=0.19$ and $\mu_B/M=3.1$. The horizon is at $r=r_H=1$.}\label{fig:exampleBackground}
\end{center}
\end{figure}

We note that the solutions found in \cite{Herzog:2009gd} are a subset of the family described by the Ansatz above with
\be
\lambda=0 \ ,  \ \Phi_M=\Phi_R=0 \ .
\ee
This fixes the coupling to the scalar $\Delta=2$ operator to zero  $M=0$, implying that conformal invariance remains unbroken. The $R$-charge and chemical potential are also zero for these solutions. When searching for black brane solutions dual to $M\neq 0$ deformations we have started with the $M=0$ solutions, introducing a small $M$ (in units of temperature or chemical potential) and then make the mass incrementally bigger. We have checked that our $M=0$ solutions reproduce the results of \cite{Herzog:2009gd}. A particular example solution is shown in Fig.~\ref{fig:exampleBackground}.

\subsection{Thermodynamics}

The temperature, $T$, and entropy density, $s$, in the dual field theory can be identified with the Hawking temperature and Bekenstein-Hawking entropy of the black brane, respectively. They are straightforward to compute from the metric close to the horizon, the first as the inverse of the period of the Euclidean time direction and the second as the area of the black brane in Planck units
\be
 T = \frac{e^{-\frac{w(r_H)}{2}}}{4\pi}g'(r_H) \ , \ s = \frac{r_H^3}{4G_5}\ .
\ee
Other thermodynamic quantities such as the energy density $\varepsilon$, pressure $p$, and baryon and $R$-charge densities $Q_R$, $Q_B$, respectively, are computed as expectation values of the energy-momentum tensor and the corresponding currents
\be
\vev{T_{00}}=\varepsilon \ , \ \vev{T_{ij}}=p \delta_{ij} \ , \ \vev{J^0_B}=Q_B \ , \ \vev{J_R^0}=Q_R \ .
\ee
The details of this calculation using holographic renormalization are relegated to Appendix~\ref{app:holoren}. 

Some of these quantities can be computed at any radius in the bulk. From the vector and Einstein equations of motion one can identify several functions that remain constant along the radial direction
\be
\partial_r \mathcal{Q}_B=\partial_r \mathcal{Q}_R=\partial_r \mathcal{H}=0 \ ,
\ee
where $\mathcal{Q}_B$ and $\mathcal{Q}_R$ where introduced before in \eqref{eq:QBQR} and
\begin{equation}
\begin{split}
 \mathcal{H} =& \frac{1}{16\pi G_5} \left( r^3 e^{-\frac{w(r)}{2}}g'(r) - r^3 e^{-\frac{w(r)}{2}}g(r)w'(r) - 2r^2 e^{-\frac{w(r)}{2}}g(r) \right) \\
 &- \Phi(r) \mathcal{Q}_B - \Phi_R(r)\mathcal{Q}_R - \Phi_M(r)\mathcal{Q}_M \ .
 \end{split}
\end{equation}
Here we have also made use of
\begin{equation}
 \mathcal{Q}_M = \frac{e^{\frac{w(r)}{2}+2\eta(r)-\frac{4\chi(r)}{3}}r^3}{16\pi G_5}\left[-\sinh(2\lambda(r))\Phi'(r) + \cosh(2\lambda(r))\left(\Phi_M'(r) - \Phi_R'(r)\right)\right] \ ,
\end{equation}
which can be thought of as the quantity that would be conserved {\emph{if}} the field $\Phi_M$ was massless (which it is not). When these quantities are evaluated at the boundary they coincide with the charges and a combination of thermodynamic potentials
\begin{equation}
\begin{split}
& \lim_{r\to \infty} \mathcal{Q}_B =Q_B\ ,\ \ \lim_{r\to \infty}  \mathcal{Q}_R = Q_R \\
&   \lim_{r\to \infty}  \mathcal{H} = \varepsilon+p-\mu_B Q_B-\mu_R Q_R\ .
 \end{split}
\end{equation}
When they are evaluated at the horizon and equated with the boundary values, one obtains expressions for $Q_B$ and $Q_R$ in terms of the fields at the horizon yielding the thermodynamic relation
\begin{equation}\label{eq:funda}
 \varepsilon + p = T s + \mu_B Q_B + \mu_R Q_R \ .
\end{equation}
Therefore, the enthalpy $\varepsilon+p$ can also be expressed in terms of fields evaluated at the horizon. 
We emphasize that the solutions to the equations of motion are rather involved, so having established the equality (\ref{eq:funda}) on the solutions is far from fortuitous and should be viewed as a highly non-trivial check of our analysis. 

In the family of solutions constructed in \cite{Herzog:2009gd} the scalar field dual to the $\Delta=2$ operator is set to zero $\lambda=0$ and the $R$-charge density of the dual vanishes $\mathcal{Q}_R=0$. The expression for the function associated to the baryonic charge in this simpler case is
\be
 \mathcal{Q}_B = e^{\frac{w}{2}+2\eta-\frac{4\chi}{3}}r^3\Phi'\  , \ M = 0 \ .
\ee
We generalize these results by allowing for a breaking of conformal invariance, $\lambda\neq 0$. This necessarily turns on the $R$-charge gauge field $\Phi_R$, forcing us to choose an ensemble --- the typical choices being grand canonical (fixed $\mu_R$) or canonical (fixed $\mathcal{Q}_R$). We elect to work in the canonical, and we set $\mathcal{Q}_R=0$ throughout. Thus we are focusing on a two-dimensional slice of the full three-dimensional phase diagram. The advantage to setting $\mathcal{Q}_R=0$ is two-fold: it simplifies the equations of motion somewhat, and in Sec.~\ref{sec:color} it will allow us to use arguments from \cite{Henriksson:2019zph} to set the probe brane angular momenta to zero. Since we also fix $\mu_B$, we are working in a mixed ensemble of fixed $R$-charge and baryon chemical potential. Note that the $R$-charge chemical potential is in general non-zero, as it cannot be tuned independently.

\begin{figure}
 \begin{center}
  \includegraphics[scale=0.8]{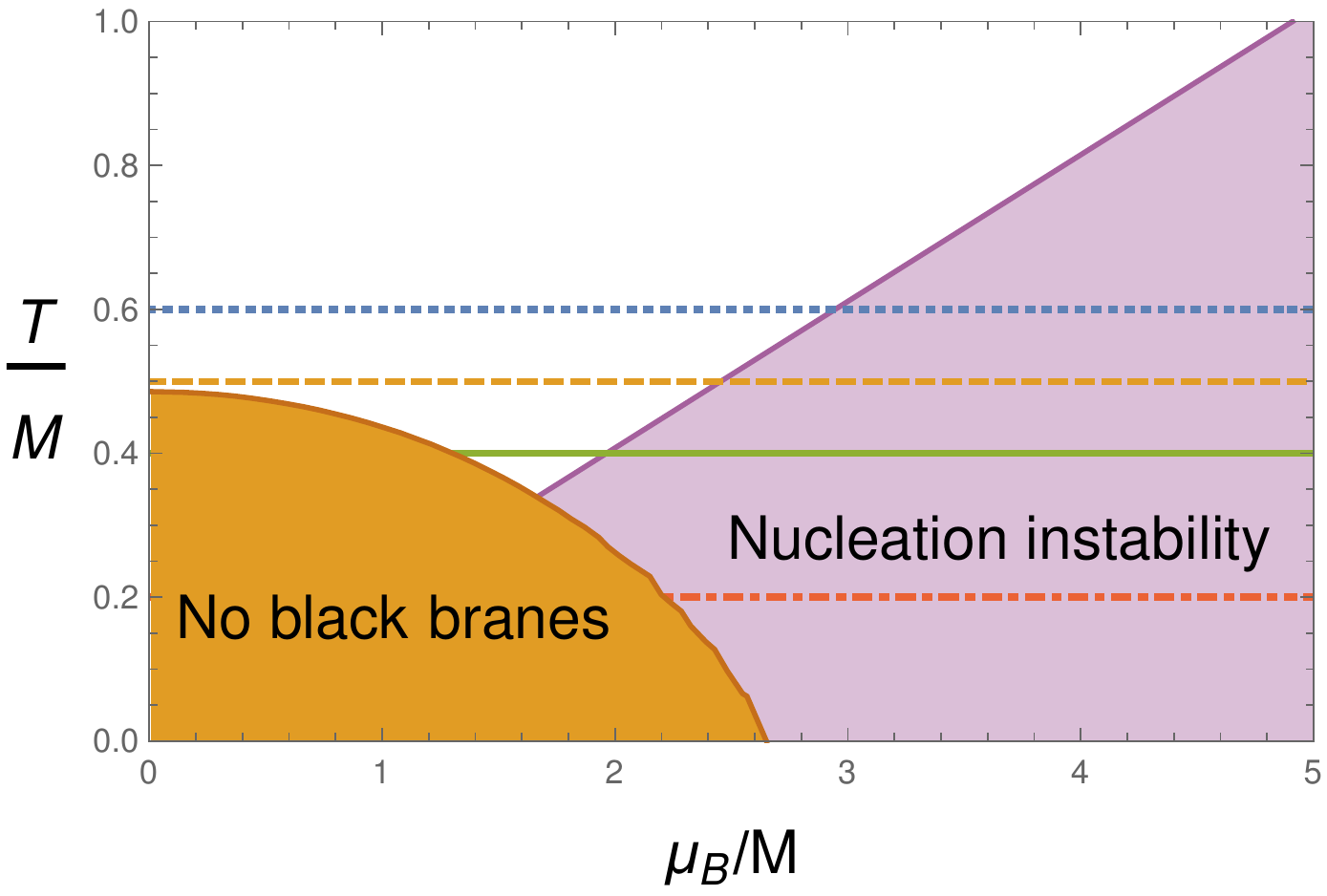}
  \caption{The phase diagram. Within the orange region we find no black brane solutions. Within the purple region the dominant black brane solutions are unstable to brane nucleation. The four horizontal lines correspond to the curves in Fig. \ref{fig:cs}.}\label{fig:phaseDiagram}
 \end{center}
\end{figure}

\begin{figure}
\begin{center}
\includegraphics[scale=0.58,valign=t]{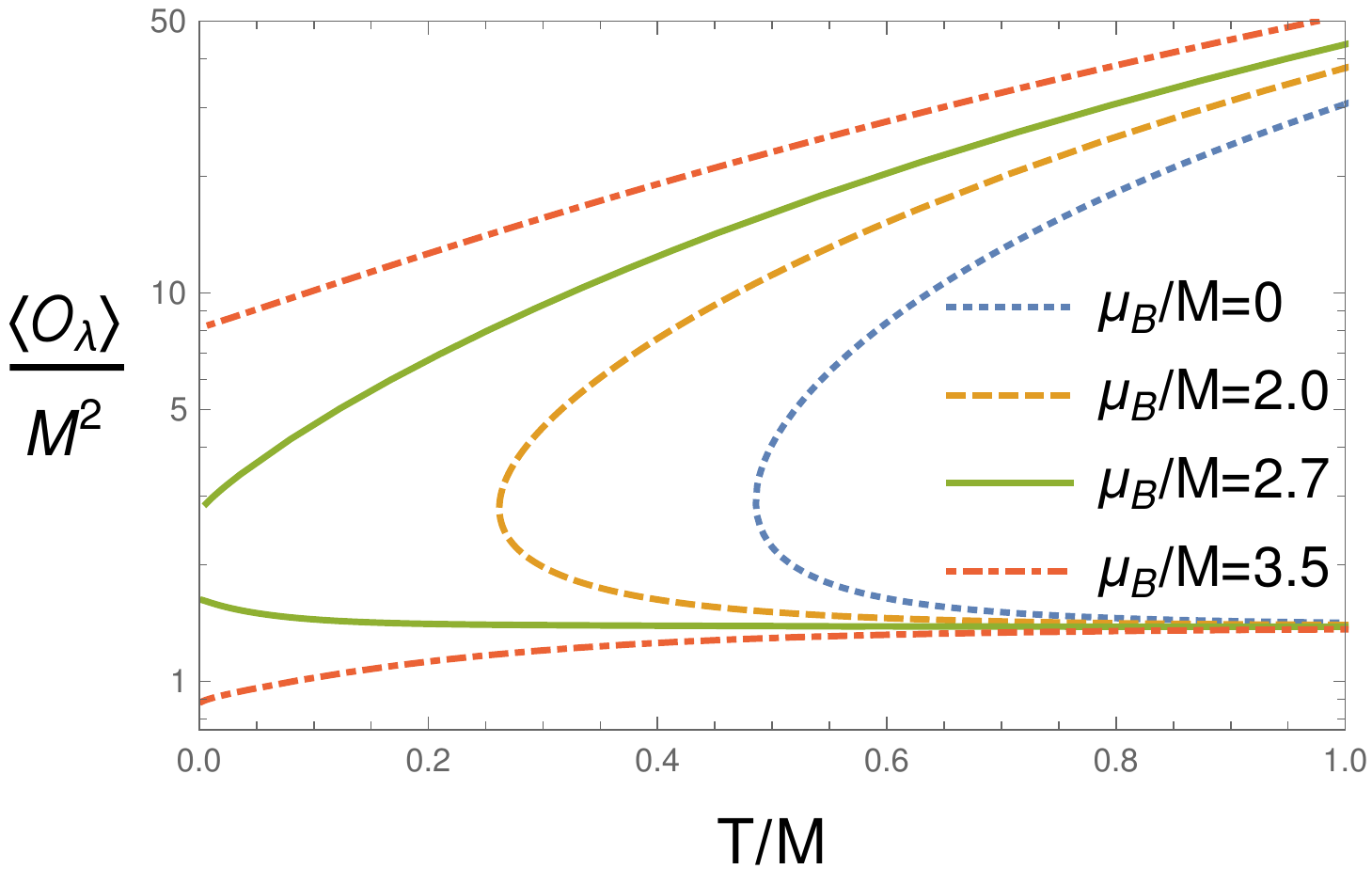}
\includegraphics[scale=0.58,valign=t]{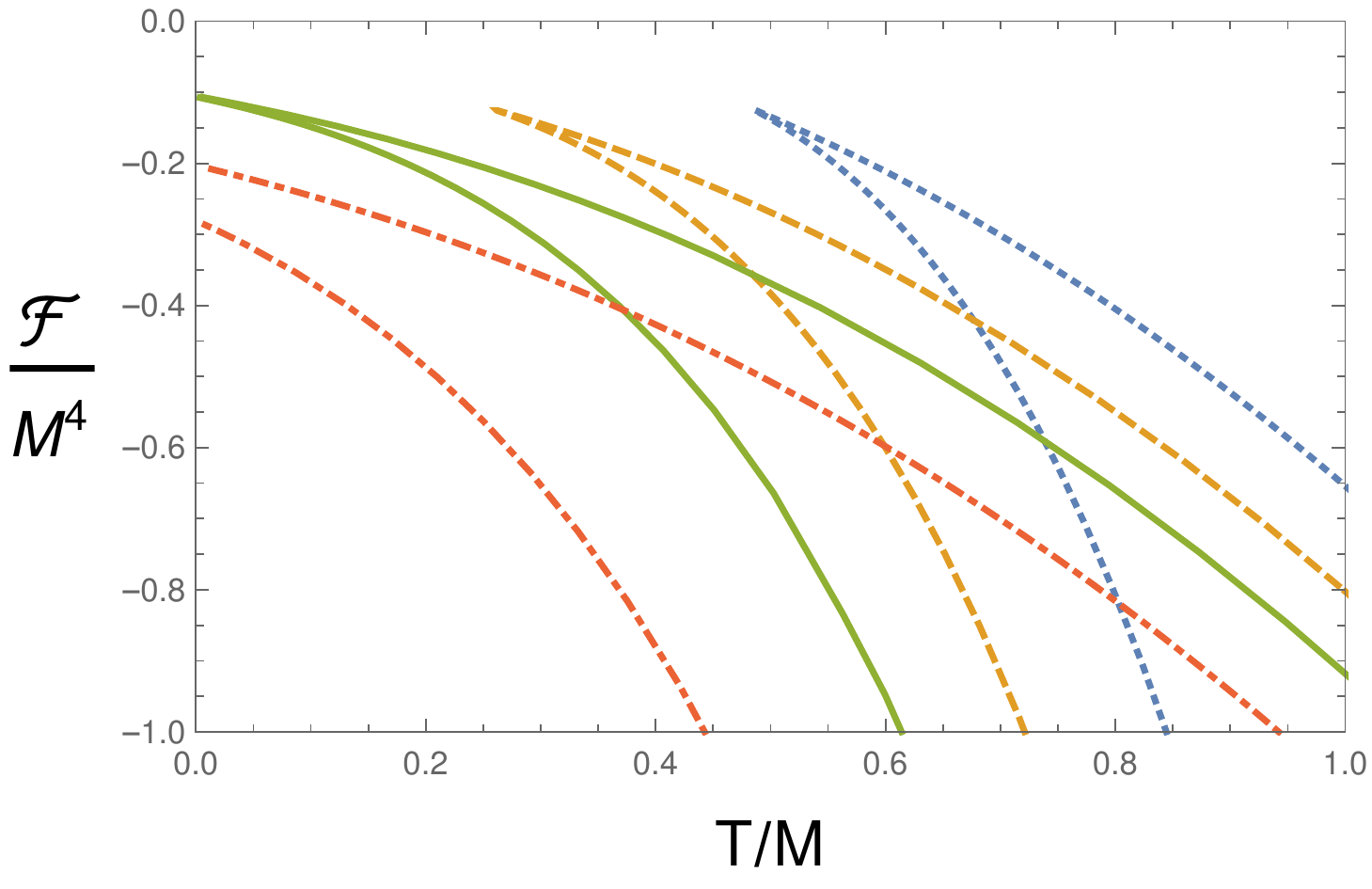}
\caption{Expectation value of the operator dual to $\lambda$ (Left panel; note the logarithmic vertical axis) and the free energy (Right panel) as functions of the temperature, at various chemical potentials. The branch of solutions with lower expectation value on the left corresponds to the one with lower free energy on the right, and is thus dominant. As the orange region in Fig.~\ref{fig:phaseDiagram} is approached, the two branches merge.}\label{fig:splittingBranches}
\end{center}
\end{figure}

We show the resulting phase diagram in the temperature and baryon chemical potential plane in Fig.~\ref{fig:phaseDiagram}. In general, we find that there are two distinct black brane solutions at each point of the diagram. As we move towards low values of temperature and chemical potential, these two branches approach each other and finally merge. Beyond the point of merging, we find no black brane solutions at all --- this is the orange region in Fig.~\ref{fig:phaseDiagram}. In Fig.~\ref{fig:splittingBranches}, the left panel shows the expectation value of the scalar dual to $\lambda$. The two branches, and their merging at low temperature and chemical potential, can be seen. Note that the branch with smaller expectation values is the one which at large temperatures and chemical potentials connects with the solutions of \cite{Herzog:2009gd}. This branch is also the one that always has the lower free energy (see the right panel of Fig.~\ref{fig:splittingBranches}), and will thus dominate the phase diagram. In the rest of the paper, we will therefore focus on this branch.

At $\mu_B=\mu_R=0$, a natural interpretation of the lack of black brane solutions at low temperatures would be that the temperature at the boundary of the orange region corresponds to the critical temperature where the effective potential \eqref{eq:VmT} becomes unbounded from below. As the chemical potential is increased one would then expect the instability to grow worse. This is however not reflected in the classical gravity solution, as the phase diagram boundary moves to lower temperatures. However, we will see that the brane nucleation instability is present in all the region of lower temperatures and that the boundary of the unstable region in the phase diagram moves to higher temperatures as the chemical potential is increased. We have depicted the unstable region in purple in Fig.~\ref{fig:phaseDiagram}.

\begin{figure}
\begin{center}
\includegraphics[scale=0.8]{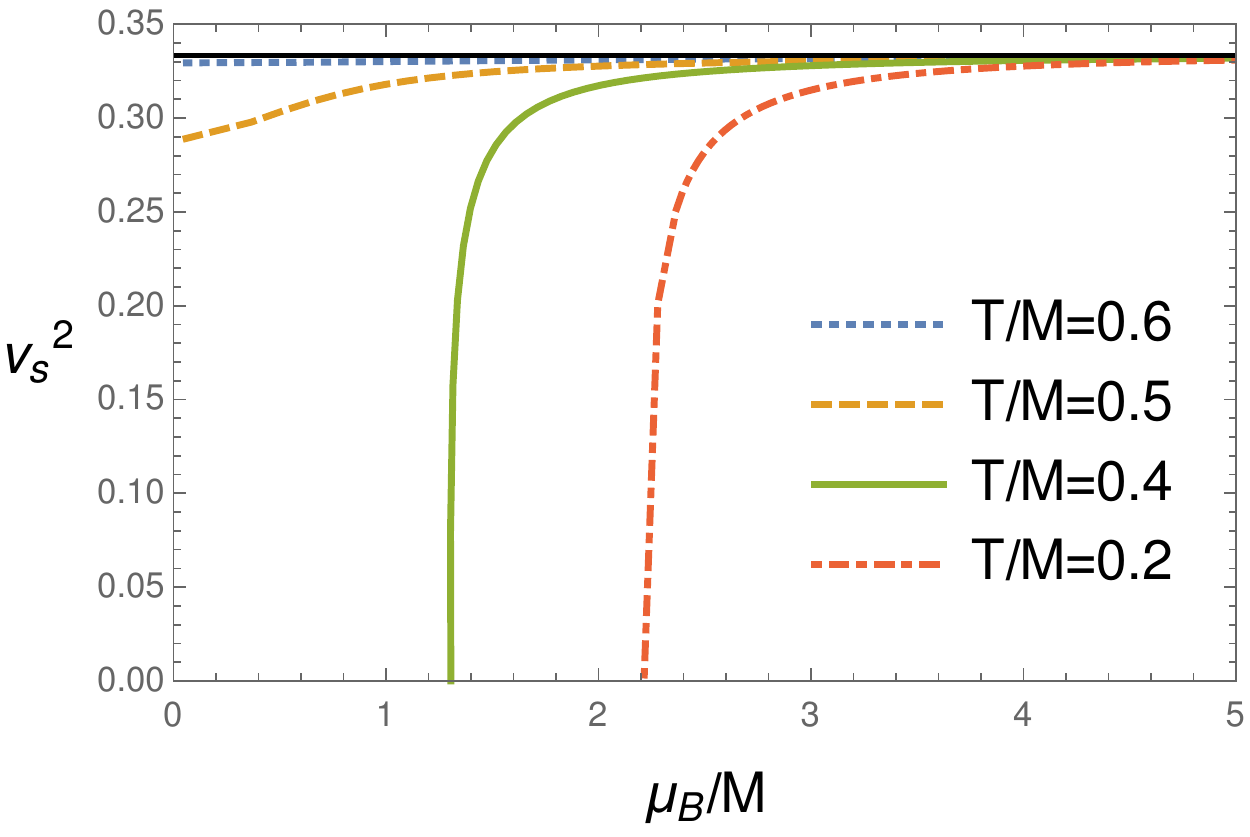}
\caption{Speed of sound squared. The horizontal black line corresponds to the conformal value of $1/3$. We observe that the speed of sound decreases rapidly to zero when the boundary of the phase diagram is approached.}\label{fig:cs}
\end{center}
\end{figure}

We have also computed the stiffness determined by the thermodynamic derivative
\be
v_s^2=\left(\frac{\partial p}{\partial \varepsilon}\right)_s \ .
\ee
At zero charge, the stiffness equals the speed of sound squared, so the system is expected to become thermodynamically unstable if $v_s^2<0$, or be inconsistent with causality if $v_s^2>1$. One can thus use the value of $v_s^2$ as a diagnostic of thermodynamic stability. In Fig.~\ref{fig:cs}, we display the results for the branch of solutions with lower free energy. We observe that $v_s^2$ is always below the conformal value and decreases significantly until it reaches zero as the boundary of the phase diagram is approached. Very near this boundary the solutions are therefore thermodynamically unstable.

\section{Brane nucleation and color superconductivity}\label{sec:color}

String theory of course allows for phenomena not captured by pure supergravity alone. In particular, holographic states can exhibit instabilities mediated by stringy processes. In this section, we will search for such instabilities. We recall that our gravitational backgrounds are sourced by a stack of $N_c$ ``color'' branes as well as a density of wrapped, ``baryonic'' branes. Following \cite{Herzog:2009gd,Henriksson:2019zph}, we will compute an effective potential for a probe brane of each type, as well as that of a D5-brane, as a function of the radial coordinate. If this potential has a global minimum \emph{outside} the horizon, we interpret this as a sign of an instability --- the branes want to condense outside of the black brane. In this section we will concentrate on the condensation of color branes, while the effective analysis for the baryonic branes follows in Sec.~\ref{sec:baryon}. To gain some geometric intuition of the brane embeddings, we have produced the table below. A dot means that the brane is localized\footnote{We have picked a co-rotating frame relative to the background geometry: In the Ansatz we write down for the probes the embedding rotates in the angular direction $\psi\sim \omega t$.} and a crossed circle that it is extended along the corresponding direction. The slashed circle on several of the internal directions corresponds to the branes wrapping a diagonal two-cycle.

\beq
\begin{array}{rccccccccccl}
 &t & x & y & z & r & \theta_1 &\phi_1  &\theta_2 &\phi_2 & \psi & \nonumber \\
N_c\text{ color D3 (background)}: &\otimes & \otimes &\otimes &\otimes &\cdot &\cdot & \cdot &\cdot &\cdot &\cdot &   \\
\text{Color probe D3}: &\otimes &\otimes &\otimes &\otimes &\cdot & \cdot & \cdot &\cdot &\cdot &\cdot & \\
\text{Probe D5}: &\otimes &\otimes & \otimes &\otimes &\cdot &\oslash &\oslash &\oslash &\oslash &\cdot & \\
\end{array}
\eeq

When discussing the embedding of the branes, we will denote their spacetime coordinates by capital letters $X^\mu$. The timelike coordinate on the brane worldvolume will always be denoted by $\tau$ --- we will sometimes choose a gauge where this equals the proper time of a certain observer on the brane, and sometimes where it equals the brane's 10D spacetime time-coordinate $\T$. Derivatives with respect to $\tau$ will be denoted by a dot throughout. Components of the 10D metric are denoted by $g_{\mu\nu}$; the metric components $g_{x_i x_i}$ with $i=\{1,2,3\}$ are all equal and are collectively denoted by $g_{xx}$.

\subsection{D3-branes}

A global minimum for the effective potential outside of the horizon signals an instability. This leads to the ``brane nucleation'' process and corresponds to what the authors of \cite{Hartnoll:2009ns} call ``Fermi seasickness''. Since these branes are of the same type as those who furnish the field theory, this instability would lead to a Higgsing of the gauge group, {\emph{i.e.}}, there is a spontaneous breaking of the gauge symmetry as $SU(N_c+1)\times SU(N_c+1)\to SU(N_c)\times SU(N_c)\times U(1)$. We interpret this as analogous to color superconductivity \cite{Rajagopal:2000wf}.

Let us compute the on-shell D3-brane action. To achieve this goal we need to be scrupulous on how to localize the probe brane in the internal directions. As can be seen from the 10D metric (\ref{eq:10Dmetric}), a non-zero $\Phi_R$ actually means that the black brane is rotating -- there are off-diagonal time-angle components of the metric, much like in the Kerr black hole in the standard Boyer-Lindquist coordinates. This is typical for $R$-charged black branes -- from the 10D point-of-view, the $R$-charge corresponds to angular momentum. Moreover, if $\Phi_R$ asymptotes to a non-zero constant, corresponding to a non-zero chemical potential for the $R$-charge, then the coordinates are rotating even at the asymptotically AdS (aAdS) boundary. We refer the reader to an exposition of the brane nucleation in a clean, analytic, framework of \cite{Henriksson:2019zph} to gain better intuition on the relevant physics.

Thus, we need to let the brane to rotate in the $\psi$ coordinate of (\ref{eq:10Dmetric}).\footnote{One could be slightly more general and consider the brane revolving in the $(\phi_1,\phi_2)$-plane. However, it is possible to show that this leads to an increase of the potential energy of the brane, making it less interesting when searching for potential minima.} To implement this, let us parametrize the worldvolume of the brane by coordinates $\xi^\alpha=(\tau,\chi_1,\chi_2,\chi_3)\in (-\infty,\infty)$. We then make the following Ansatz for the embedding:
\begin{equation}\label{eq:probecoords}
 \begin{gathered}
  \T=\T(\tau) \, , \qquad \R=\R(\tau) \, , \qquad X_i=\chi_i \quad (i=1,2,3) \, , \\
  \Theta_1 = \theta_1^0 \, , \qquad \Phi_1 = \phi_1^0 \, , \qquad \Theta_2 = \theta_2^0 \, , \qquad \Phi_2 = \phi_2^0 \, , \qquad \Psi=\Psi(\tau) \ .
 \end{gathered}
\end{equation}
In (\ref{eq:probecoords}) the quantities with superscript ``0'' are constants specifying the location of the D3-brane in the internal directions -- these will not enter into the final result. The action of the D3-brane reads
\begin{equation}\label{eq:coloraction}
 S_{D3} = -T_3 \int d^4\xi\sqrt{-\det(P[g_{\mu\nu}])} + T_3 \epsilon_3 \int P[C_4] \equiv \int d^4\xi\, \mathcal{L}_{D3} \ ,
\end{equation}
where we defined the Lagrangian density $\mathcal{L}_{D3}$. Here, $P[\cdot]$ denotes the pullback of a 10D spacetime field to the brane worldvolume, and $\epsilon_3=+1$ ($\epsilon_3=-1$) for a D3-brane ({$\overline{\rm{D3}}$}-brane). The dilaton is constant in our backgrounds, and $T_3 = (2\pi)^{-3} g_s^{-1} \alpha'^{-2}$ where $\alpha'^{-2}=4\pi g_s N_c\cdot \frac{27}{16}$. To evaluate the DBI term it is convenient to consider an observer located on the brane at fixed worldvolume coordinates $(\chi_1,\chi_2,\chi_3)$. Taking the derivative with respect to $\tau$ gives the velocity vector
\begin{equation}
 U \equiv \frac{dX^\mu}{d\tau}\partial_\mu = \dot \T\partial_t + \dot \R \partial_r + \dot\Psi \partial_{\psi} \ .
\end{equation}
The induced line element can then be written as
\begin{equation}
 ds_4^2 = U_\mu U^\mu d\tau^2 + g_{xx} \left( d\chi_1^2 + d\chi_2^2 + d\chi_3^2 \right) \ ,
\end{equation}
and the square root in the DBI term as
\begin{equation}
 \sqrt{-\det(P[g_{\mu\nu}])} = g_{xx}^{3/2}\sqrt{-U_\mu U^\mu} \ , \label{eq:D3DBIsimple}
\end{equation}
where
\begin{equation}\label{eq:D310velSquared}
\begin{split}
 U_\mu U^\mu =& g_{tt}\dot \T^2 + g_{rr}\dot \R^2 + g_{\psi\psi}\dot \Psi^2 + 2g_{t\psi}\dot \T \dot\Psi \ .
\end{split}
\end{equation}
Below, after performing the variations of the action, we fix $\tau$ to be the proper time of the observer, such that the velocity squares to minus one: $U_\mu U^\mu =-1$. Finally, the WZ term $P[C_4]$ in (\ref{eq:coloraction}) becomes
\begin{equation}
 P[C_4] = \left[ (C_4)_t \dot \T + (C_4)_{\psi} \dot\Psi \right] d\tau \wedge d\chi_1 \wedge d\chi_2 \wedge d\chi_3 \ ,
\end{equation}
where $(C_4)_t$ and $(C_4)_\psi$ denote the $(t, x_1, x_2, x_3)$-component and the $(\psi, x_1, x_2, x_3)$-component of $C_4$, respectively.

Recall that our probe D3-brane is not a static object in the ambient background metric. Rather, it is bound to geodesics with radial and angular sway. The probe D3-brane has two conserved quantities: the total energy and the angular momentum which we extract from the action. The energy and angular momentum can be determined by varying the Lagrangian density $\mathcal{L}_{D3}$ with respect to $\dot \T$ and $\dot\Psi$, respectively.  We arrive at the following expressions
\begin{align}
 E &\equiv -\frac{1}{T_3}\frac{\partial \mathcal{L}_{D3}}{\partial \dot \T} = g_{xx}^{3/2} \left( -g_{tt}\dot \T - g_{t\psi}\dot\Psi \right) - \epsilon_3 (C_4)_t \label{eq:D3energy} \\ 
 J_\psi &\equiv \frac{1}{T_3}\frac{\partial \mathcal{L}_{D3}}{\partial \dot\Psi} = g_{xx}^{3/2} \left( g_{t\psi}\dot \T + g_{\psi\psi}\dot\Psi \right) + \epsilon_3 (C_4)_{\psi} \label{eq:D3angMomPsi} \ .
\end{align}
We have simplified the result of the variations using $U_\mu U^\mu=-1$. We can now use (\ref{eq:D3energy}) and (\ref{eq:D3angMomPsi}) together with $U_\mu U^\mu =-1$ to solve for the energy in terms of the angular momentum. Since $U_\mu U^\mu$ given in (\ref{eq:D310velSquared}) is quadratic in $\dot \T$ we get two branches; we pick the one with $\dot \T>0$. Our result is
\begin{equation}
 E_{D3} = - \epsilon_3 (C_4)_t - g_{t\psi}\frac{J_\psi - \epsilon_3 (C_4)_{\psi}}{g_{\psi\psi}} + \sqrt{\left( -g_{tt} + \frac{g_{t\psi}^2}{g_{\psi\psi}} \right)\left(g_{xx}^3 \left( 1+g_{rr}\dot \R^2 \right) + \frac{\left( J_\psi - \epsilon_3(C_4)_{\psi} \right)^2}{g_{\psi\psi}}  \right)} \ .
\end{equation}
The effective potential is defined to be the energy with $\dot \R=0$:
\begin{equation}
 \begin{split}
 V_{D3} = - \epsilon_3 (C_4)_t - g_{t\psi}\frac{J_\psi - \epsilon_3 (C_4)_{\psi}}{g_{\psi\psi}} + \sqrt{\left( -g_{tt} + \frac{g_{t\psi}^2}{g_{\psi\psi}} \right)\left(g_{xx}^3 + \frac{\left( J_\psi - \epsilon_3(C_4)_{\psi} \right)^2}{g_{\psi\psi}}  \right)} \ .
 \end{split}
\end{equation}
We note that this expression only depends on $J_\psi$ and the radial position. Plugging in the explicit metric components, we arrive at
\begin{equation}
 V_{D3} = -\epsilon_3 (C_4)_t - \frac{3\Phi_R}{\sqrt{2}}\left(J_\psi - \epsilon_3 (C_4)_\psi \right) + e^{-\frac{w}{2}-\frac{10}{3}\chi}\sqrt{g} \sqrt{ r^6 + 9 e^{4\eta+4\chi} \left(J_{\psi} - \epsilon_3 (C_4)_{\psi} \right)^2 } \ . \label{eq:D3effPotSimple}
\end{equation}
Expanding this for large radii we find
\begin{equation}
 V_{D3} = \left( 1-\epsilon_3 \right)r^4 + \mathcal{O}(r^{2}) \ .
\end{equation}
This result confirms that $\epsilon_3=-1$ corresponds to a brane of opposite charge to the ones sourcing the background --- such a {$\overline{\rm{D3}}$}-brane is always attracted towards the horizon at large radii. By also plotting the full potential for various backgrounds, we find that the force on {$\overline{\rm{D3}}$}-branes is directed toward the horizon for all values of the radial coordinate, in all of the available phase diagram.

On the other hand, $\epsilon_3=+1$ corresponds to a D3-brane of the same type as those sourcing the background. In this case, one can see that the potential instead approaches a constant value:
\begin{equation}
 V_{D3} = C - \frac{3\mu_R}{\sqrt{2}} J_\psi + \mathcal{O}(r^{-1}) \ .
\end{equation}
The overall sign of this depends on the parameters of the background, in addition to the angular momentum of the probe $J_\psi$.\footnote{Explicitly, $C$ can be written in terms of coefficients of the boundary expansion \eqref{eq:UVexpansion} as $C=g_{2,0}/2-\lambda_{2,0}^2/6+\lambda_{2,0}\lambda_{2,1}/8-5\lambda_{2,1}^2/192-\Phi_{R\,2,0}\Phi_{R\, 0,0}$.} As was argued in \cite{Henriksson:2019zph}, the magnitude of the angular momentum is given by the average angular momentum of the branes that source the background (other values are statistically suppressed). Since our backgrounds all have zero angular momentum (zero $R$-charge), we will set $J_\psi=0$.

\begin{figure}
\begin{center}
\includegraphics[scale=0.75]{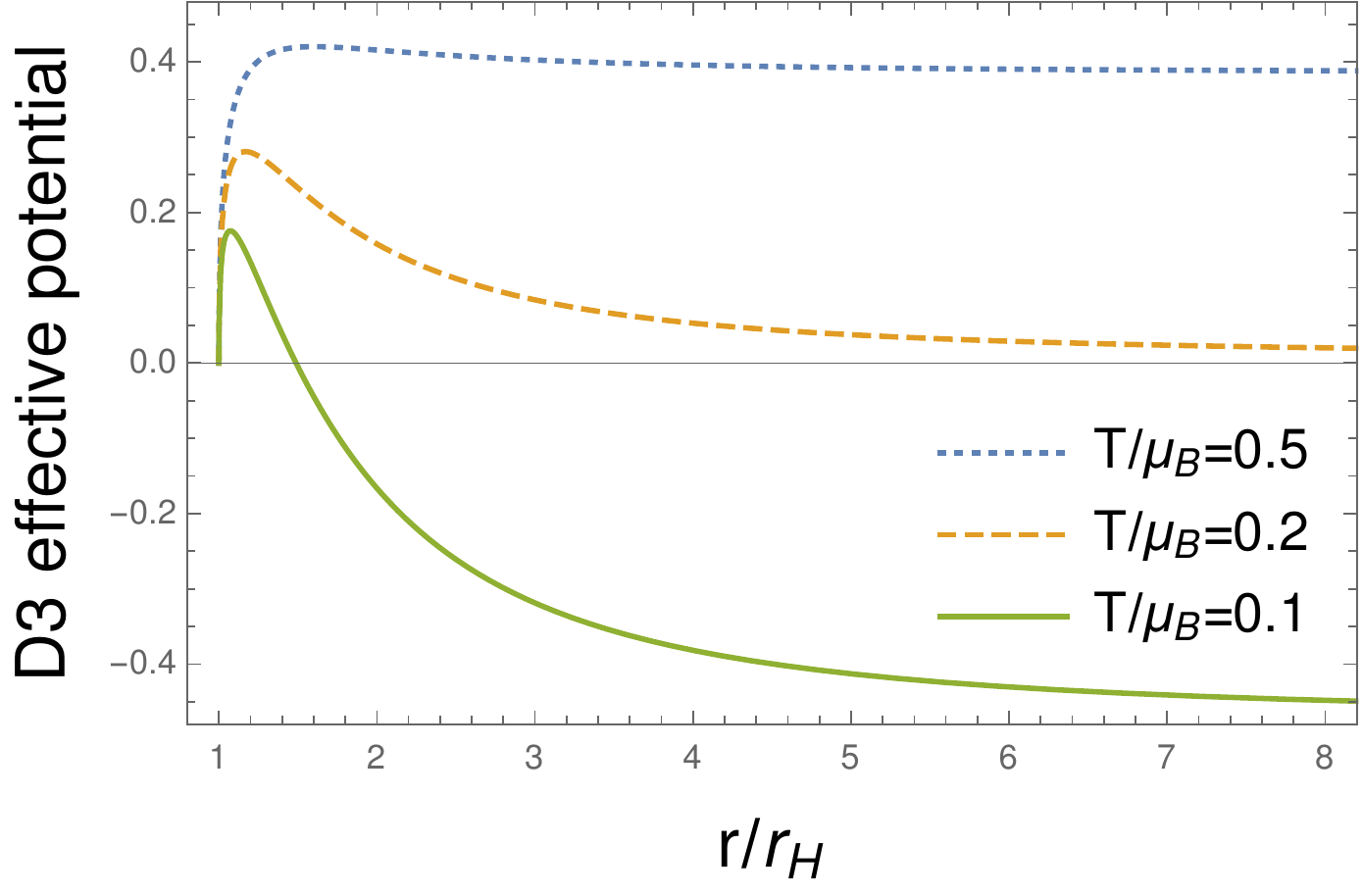}
\caption{The effective potential for a probe color D3-brane at $T/M=0.5$, with $J_\psi=0$.}\label{fig:colorD3effPot}
\end{center}
\end{figure}

In Fig.~\ref{fig:colorD3effPot} we have depicted the effective potential to illustrate that at low temperatures relative to the baryon chemical potential, the asymptotic value for the potential dives below zero, which is the value at the horizon. This suggests that the D3-branes which are cloaked by the event horizon can lower their energy by tunneling through the potential barrier and moving away towards the boundary of spacetime. In doing so, the gauge group is Higgsed as was discussed above. We find that this nucleation instability occurs for $T\leq 0.20\mu_B$, in agreement with \cite{Herzog:2009gd} for the special case of zero source for the operator dual to $\lambda$. 
In fact, our numerical analysis suggests that $T/\mu_B\approx 0.20$ for all $M/\mu_B$, implying that the on-set of the instability is insensitive to the conformal symmetry breaking of the type we are considering.
The region for the instability is displayed in the phase diagram of Fig.~\ref{fig:phaseDiagram}.

\subsection{D5-branes}

The backgrounds we are probing are built up out of only D3-branes, so one might ask why studying D5-branes is interesting. D5-branes actually play an interesting role on the conifold. The base space $T^{1,1}$ is topologically $S^2\times S^3$. It is natural then to wrap the D5-brane on the $S^2$ and extend it in the field theory dimensions. This forms a domain wall in the radial direction\footnote{Such a wrapped D5-brane is also related to fractional D3-branes. These are D5-branes pinned to the conifold singularity.} and leads to a step in the rank of \emph{one} of the gauge groups \cite{Gubser:1998fp}: on one side, $SU(N_c)\times SU(N_c)$, on the other $SU(N_c)\times SU(N_c-1)$. A configuration with a D5-brane of this type outside the horizon thus describes the Higgsing of one of the gauge groups. In order to keep the fluxes of the theory unchanged, an $\overline{\rm{D5}}$-brane should also be present, so this configuration is not reached directly by the emission of color branes by the black brane but by the formation of a pair of five-branes that become separated in the bulk. Note that supersymmetry is broken when both branes and anti-branes are present, but this is natural as temperature and chemical potentials are already breaking it. By studying the effective potential of the five-branes we can determine if the background becomes unstable due to pair production or, from the field theory point of view, if asymmetric Higgsing is possible.

We now write down our Ansatz for the embedding. Following \cite{Dasgupta:1999wx,Arean:2004mm}, we consider the D5-brane to wrap an $S^2$ of the conifold, schematically $S^2_1 - S^2_2$ where $S^2_1$ and $S^2_2$ are the two-spheres furnishing the base of the conifold. To accomplish this, we introduce brane worldvolume coordinates $(\tau,\chi_1,\chi_2,\chi_3,\alpha,\beta)$, where $0\leq\alpha<\pi$ and $0\leq\beta<2\pi$ are angular coordinates on the $S^2$, while the other coordinates take values from $-\infty$ to $\infty$. Our Ansatz for the embedding is then as follows:
\begin{equation}
 \begin{gathered}
  \T=\T(\tau) \, , \qquad \R=\R(\tau) \, , \qquad X_i=\chi_i \quad (i=1,2,3) \, , \\
  \Theta_1 = \alpha \, , \qquad \Phi_1 = \beta \, , \qquad \Theta_2 = \alpha \, , \qquad \Phi_2 = -\beta \, , \qquad \Psi=\Psi(\tau) \ .
 \end{gathered}
\end{equation}
Notice that besides the radial motion, the brane has an angular velocity $\dot\Psi$. Note also that we could had considered letting the D5-brane rotate in the $\phi_i$-directions, however, as was the case for the extended D3-branes, this will only add a positive contribution to the effective potential and is thus not very interesting.

We have learned that the primary reason for the brane nucleation to occur is due to having non-trivial WZ terms in the action. The only form-field turned on in our backgrounds is $C_4$. Thus, to obtain a non-zero WZ part of the D5-brane action we seek to turn on a worldvolume gauge field ${\mathcal{F}}$. This can be simply done by turning on a magnetic flux on the $S^2$ that the D5-brane wraps. The action for the D5-brane is then
\begin{equation}
 S_{D5} = -T_5 \int d^6\xi\sqrt{-\det(P[g_{\mu\nu}]+\mathcal{F})} + T_5\, \epsilon_5 \int P[C_4]\, \wedge\, \mathcal{F} \ ,
\end{equation}
where $T_5 = (2\pi)^{-5} g_s^{-1} \alpha'^{-3}$, and $\epsilon_5=+1$ ($\epsilon_5=-1$) corresponds to a D5 ($\overline{\rm{D5}}$). We turn on the following worldvolume flux:
\begin{equation}
 \mathcal{F} = f \sin{\alpha}\ d\alpha\wedge d\beta \ .
\end{equation}
This gives a D3-brane charge to the D5-brane --- effectively we dissolve a number of D3-branes in the D5-brane. For a D5-brane ($\epsilon_5=+1$), a positive flux corresponds to dissolving D3-branes while a negative flux corresponds to dissolving $\overline{\rm{D3}}$-branes. For an $\overline{\rm{D5}}$-brane ($\epsilon_5=-1$), the opposite is true. The flux $f$ is in fact quantized, $f=\pi\alpha' n$ for integer $n$ \cite{Arean:2006vg}. However, since $\alpha'\sim 1/\sqrt{g_s N_c}$, we can regard it as a continuous parameter in the limit we are working in. Dissolving exactly one D3-brane would correspond to choosing $n=1$. Below we consider fluxes of order 1 and are therefore dissolving a large number $\sqrt{g_s N_c}$ of D3-branes (though not larger than what is allowed by the probe limit).

From the previous subsection we know that adding $\overline{\rm{D3}}$'s should add to the attractive force between the D5-brane and the stack of background D3-branes. Dissolving D3-branes should on the other hand add a repulsive component to the force in the low-temperature region of the phase diagram. We therefore expect that dissolving a sufficient number of D3-brane charge in the D5-brane will give rise to an instability in the effective potential.

To check if this is borne out, we proceed along similar lines as in the previous subsection. We consider an observer located on the D5-brane at fixed worldvolume coordinates $(\chi_1,\chi_2,\chi_3,\alpha,\beta)$. Taking the derivative with respect to $\tau$, yields the velocity
\begin{equation}
 U \equiv \frac{dX^\mu}{d\tau}\partial_\mu = \dot \T\partial_t + \dot \R \partial_r + \dot\Psi \partial_{\psi} \ ,
\end{equation}
for which
\begin{equation}\label{eq:D510velSquared}
\begin{split}
 U_\mu U^\mu =& g_{tt}\dot \T^2 + g_{rr}\dot \R^2 + g_{\psi\psi}\dot\Psi^2 + 2g_{t\psi}\dot \T \dot\Psi \ .
\end{split}
\end{equation}
This velocity squares to minus one, $U_\mu U^\mu = -1$, upon fixing $\tau$ to the proper time. We can then write the induced line element on the brane worldvolume as
\begin{equation}
 ds_6^2 = U_\mu U^\mu d\tau^2 + g_{xx} \left( d\chi_1^2 + d\chi_2^2 + d\chi_3^2 \right) + \left( g_{\theta_1\theta_1} + g_{\theta_2\theta_2} \right) d\alpha^2 + \left(g_{\phi_1\phi_1}-2g_{\phi_1\phi_2}+g_{\phi_2\phi_2}\right) d\beta^2 \ ,
\end{equation}
and the square root in the DBI term as
\begin{equation}
 \sqrt{-\det(P[g_{\mu\nu}]+\mathcal{F})} = \Upsilon \sqrt{-U_\mu U^\mu} \ , \label{eq:DBIsimple}
\end{equation}
where we defined the quantity
\begin{equation}
 \Upsilon \equiv g_{xx}^{3/2}\sqrt{\left(g_{\theta_1\theta_1} + g_{\theta_2\theta_2}\right)\left(g_{\phi_1\phi_1}-2g_{\phi_1\phi_2}+g_{\phi_2\phi_2}\right) + f^2\sin^2 \alpha} \ .
\end{equation}
The non-vanishing components of the $P[C_4]$ are
\begin{equation}
 P[C_4] = \left[ (C_4)_t \dot \T + (C_4)_{\psi} \dot\Psi \right] d\tau \wedge d\chi_1 \wedge d\chi_2 \wedge d\chi_3 
\end{equation}
building up the WZ term
\begin{equation}
 T_5\, \epsilon_5 \int d^6\xi \left\{ \left[ (C_4)_t \dot \T + (C_4)_{\psi} \dot\Psi \right] f\sin{\alpha} \right\} \ .
\end{equation}
Above, $(C_4)_t$ and $(C_4)_\psi$ denote the $(t, x_1, x_2, x_3)$- and $(\psi, x_1, x_2, x_3)$-components of $C_4$, respectively.

\begin{figure}[h!]
\begin{center}
\includegraphics[scale=0.68,clip=true,trim=0pt 78pt 0pt 70pt]{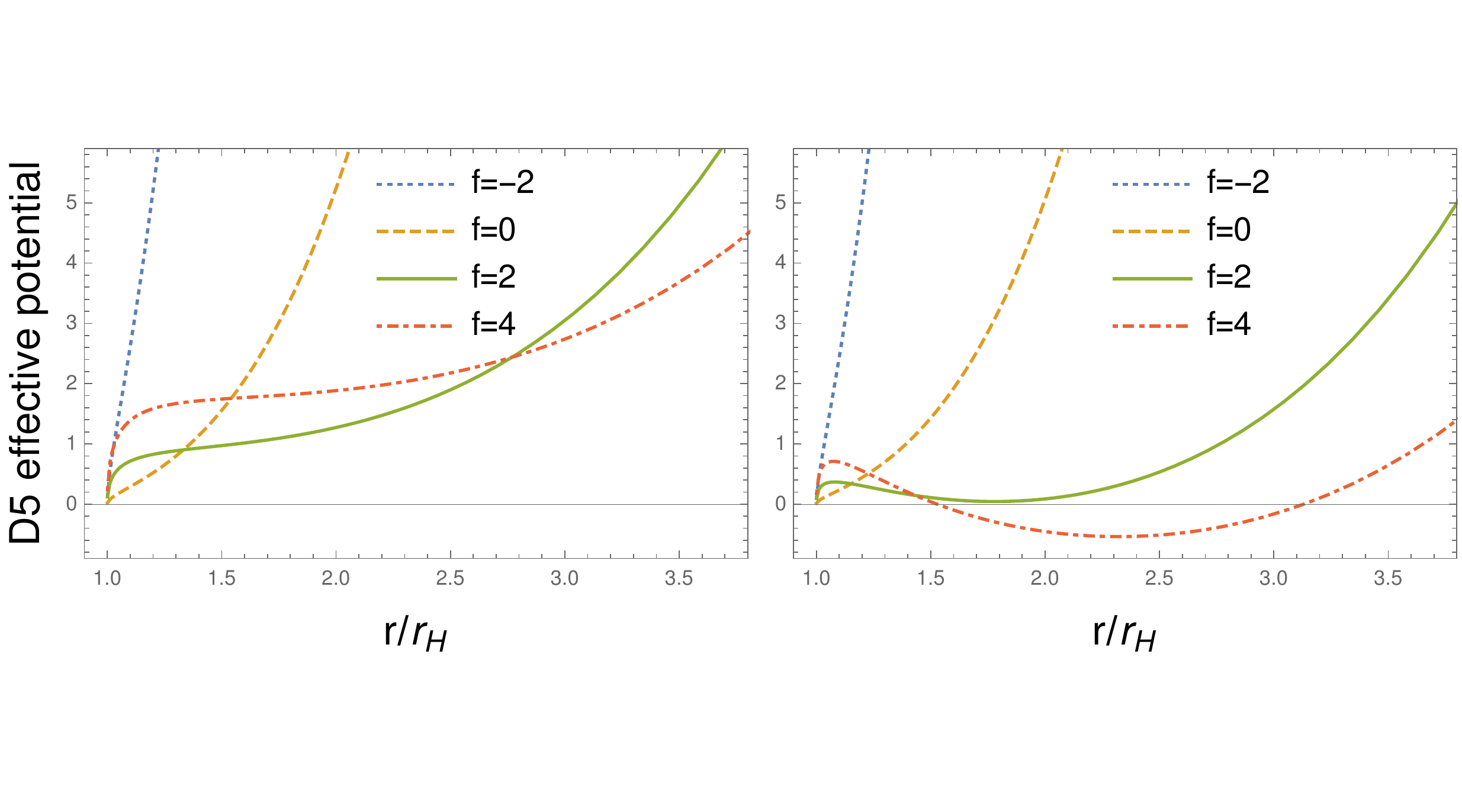}
\caption{The effective potential for a probe D5-brane, with $J_\psi=0$ and varying the worldvolume flux, in backgrounds with $T/M=0.5$. On the left, $T/\mu_B=0.5$, which is outside the region where the nucleation instability occurs --- thus no minimum forms as the flux is increased. On the right, we are inside the unstable region with $T/\mu_B=0.1$ --- increasing the flux then leads to the formation of a minimum.}\label{fig:D5effPot}
\end{center}
\end{figure}

To obtain the effective potential, we consider the two obvious conserved quantities of the resulting D5-brane action, the energy and the angular momentum. Recall that these are all really densities, since the worldvolume is infinite. The energy and the angular momentum can be determined by varying with respect to $\dot \T$ and $\dot\Psi$, respectively. We use $U_\mu U^\mu = -1$ after varying to simplify the resulting expressions. We also want to integrate over the angular coordinates $\alpha$ and $\beta$. The dependence on these coordinates is in the expression for $\Upsilon$ as well as in the WZ term, both of which are proportional to $\sin{\alpha}$. We thus define the quantity $\widetilde\Upsilon\equiv V_{S^2}^{-1}\int_{S^2}\Upsilon$, where $V_{S^2}=4\pi$. We can then write down the following conserved quantities:
\begin{align}
 E &\equiv -\frac{1}{T_5 V_{S^2}} \int_{S^2} \frac{\partial \mathcal{L}}{\partial \dot \T} = \frac{1}{V_{S^2}}\int_{S^2} \left\{ \Upsilon \left( -g_{tt}\dot \T - g_{t\psi}\dot\Psi \right) - \epsilon_5 (C_4)_t f\sin{\alpha} \right\} \nonumber \\
 &=  \widetilde\Upsilon \left( -g_{tt}\dot \T - g_{t\psi}\dot\Psi \right) - \epsilon_5 (C_4)_t f \label{eq:D5energy} \\
 J_\psi &\equiv \frac{1}{T_5 V_{S^2}} \int_{S^2} \frac{\partial \mathcal{L}}{\partial \dot\Psi} = \frac{1}{V_{S^2}}\int_{S^2}\left\{ \Upsilon \left( g_{t\psi}\dot \T + g_{\psi\psi}\dot\Psi \right) + \epsilon_5 (C_4)_\psi f\sin{\alpha} \right\} \nonumber \\
 &=  \widetilde\Upsilon \left( g_{t\psi}\dot \T + g_{\psi\psi}\dot\Psi \right) + \epsilon_5 (C_4)_\psi f\label{eq:D5angMomPsi} \ .
\end{align}
We can now use (\ref{eq:D5energy}) and (\ref{eq:D5angMomPsi}) together with $U_\mu U^\mu=-1$ to solve for the energy in terms of the angular momentum and the flux $f$. Since (\ref{eq:D510velSquared}) is quadratic in $\dot \T$ we get two branches; we pick the one with $\dot \T>0$. The result is
\begin{equation}
 E = -\epsilon_5 (C_4)_t f - g_{t\psi}\frac{J_\psi - \epsilon_5 (C_4)_\psi f }{g_{\psi\psi}} + \sqrt{\left( -g_{tt} + \frac{g_{t\psi}^2}{g_{\psi\psi}} \right)\left( \widetilde\Upsilon^2\left(1+g_{rr}\dot\R^2\right) + \frac{J_\psi - \epsilon_5 (C_4)_\psi f}{g_{\psi\psi}} \right)} \ .
\end{equation}
Again, the effective potential is defined to be the energy with $\dot \R=0$:
\begin{equation}
 V_{D5} = -\epsilon_5 (C_4)_t f - g_{t\psi}\frac{J_\psi - \epsilon_5 (C_4)_\psi f }{g_{\psi\psi}} + \sqrt{\left( -g_{tt} + \frac{g_{t\psi}^2}{g_{\psi\psi}} \right)\left( \widetilde\Upsilon^2 + \frac{J_\psi - \epsilon_5 (C_4)_\psi f}{g_{\psi\psi}} \right)} \ .
\end{equation}
After plugging in the explicit metric components, this becomes
\bea
  V_{D5} & = & -\epsilon_5 (C_4)_t f - \frac{3\Phi_R}{\sqrt{2}}\left(J_\psi - \epsilon_5 f (C_4)_\psi \right) \nonumber\\
   & &+ e^{-\frac{w}{2}}\sqrt{ 9 e^{4\eta-\frac{8}{3}\chi} g \left( J_\psi - \epsilon_5 f (C_4)_\psi \right)^2 + r^6\, g\, e^{-\frac{20}{3}\chi}\left( \frac{1}{9}e^{2(\eta+\chi)} \cosh^2(2\lambda) + f^2 \right) } \ .\label{eq:D5pot}
\eea
Finally, let us study the large radius asymptotics of the effective potential as was also done in the case of D3-brane probes. Expanding the potential (\ref{eq:D5pot}) for large radii we find
\begin{equation}
 V_{D5} = \left( \frac{1}{3}\sqrt{1+9\, \epsilon_5^2 f^2} - \epsilon_5 f \right)r^4 + \ldots \ .
\end{equation}
The quantity in parentheses is always positive, approaching zero only for $\epsilon_5 f\rightarrow\infty$. So for finite flux the D5-brane will never shoot off to the aAdS boundary. However, by making the flux large and positive we can push the region where this $r^4$ growth dominates to large radii, so depending on the subleading behavior we might get a global minimum at finite radius. Indeed this is the case, as can be seen in on the right panel in Fig.~\ref{fig:D5effPot}. More precisely, at low temperature, as the flux is increased from zero, a minimum forms at finite radius. Increasing the flux further pushes down this minimum until it dips below zero --- this signals the onset of a nucleation instability. At high temperature, increasing the flux does not lead to the formation of a minimum, however, as is shown on the left panel in Fig.~\ref{fig:D5effPot}. As one might have expected, the onset of the instability seems to occur exactly at the same $T/\mu_B$ as the D3-brane instability of the previous subsection. (Note that we have taken $J_\psi=0$ for the same reason as in the previous subsection.)

It is worthwhile to compare the potential for the D5-brane to the potential for the D3-brane worked out in the previous subsection. As argued above, the flux $f$ counts the number of D3-branes dissolved in the D5-brane. One might then anticipate that for large flux, the D5-brane potential is more and more dominated by these dissolved D3-branes. To verify this, set $\epsilon_5=+1$ in $V_{D5}$, then scale $J_\psi \rightarrow f J_\psi$ and expand for large $f$, keeping only the leading order. Finally, divide by the number of dissolved D3-branes, which is set by $f$. Doing this, we indeed recover the result (\ref{eq:D3effPotSimple}) for a single D3-brane. More precisely, for $f$ large and positive we recover (\ref{eq:D3effPotSimple}) with $\epsilon_3=1$, and for $f$ large and negative we recover (\ref{eq:D3effPotSimple}) with $\epsilon_3=-1$.


\section{Wrapped D3-branes and baryon condensation}\label{sec:baryon}

Our black brane backgrounds are charged under the baryonic $U(1)$. In finite density holography, one often encounters instabilities in charged black branes at low temperature, where the black brane prefers to shed its charge. This is the case in the famous holographic ``superconductor'', where at low temperature it is preferential for the charge to be carried by a scalar field in the bulk, rather than by the black brane. One might ask if the same happens in our geometries. A significant difference is that since no elementary field is charged under the baryonic symmetry, the objects condensing would have to be the baryonic D3-branes wrapping parts of the internal space \cite{Witten:1998xy}. Constructing the geometry resulting from such condensation will therefore be complicated. However, we can still go ahead and compute the effective potential for a probe brane, which will tell us if an instability towards condensation exists.

Before going into the details of the calculation, it is useful to recall what the dual interpretation of a bulk state with condensed D3-branes would be. In condensed matter applications of holography, the division into charge carried by black hole horizons and charge carried by matter fields outside the horizon has been discussed in depth. In the dual field theory, this is thought to correspond to the charge density being distributed over ``fractionalized'' and ``cohesive'' degrees of freedom \cite{Hartnoll:2016apf}, respectively. Here, ``fractionalized'' essentially means color-charged, {\emph{e.g.}}, quarks, while ``cohesive'' essentially means color-neutral, {\emph{e.g.}}, hadrons. This should be compared with the criteria for confinement in the bulk, which essentially depends on the existence of a horizon. If the state is confined, no horizon exists, and all charge is necessarily carried by non-dissipative cohesive excitations. However, in a deconfined state with a horizon, there are several options: The charge can be carried by cohesive excitations, by fractionalized excitations, or by a mix of the two. In the presence of fundamental fermionic matter, when all the charge is cohered, the corresponding dual field theory state can have an interpretation in terms of gapped quantum Hall states \cite{Bergman:2010gm,Jokela:2011eb}.

Let us now describe the computation for the effective potential by wrapping D3-branes on a 3-cycle in the internal space. Recall that the conifold is a $U(1)$ fiber over $S^2\times S^2$. The two simplest possibilities \cite{Gubser:1998fp} are then to wrap the $U(1)$ fiber together with one of the $S^2$'s. As pointed out in \cite{Arean:2004mm}, however, there is a larger family of embeddings, where the brane wraps the first $S^2$ an integer $m_1$ times and the second $S^2$ an integer $m_2$ times. We will see explicitly below that the winding numbers $m_1$ and $m_2$ together set the baryonic charge and the $R$-charge of the corresponding dual operator (roughly, higher winding numbers correspond to larger charges).

To implement this embedding, let us parametrize the brane worldvolume by the coordinates $(\tau,\zeta,\beta,\gamma)$, where $-\infty<\tau<\infty$, $0\leq\zeta<\infty$, $0\leq\beta<2\pi$, and $0\leq\gamma<4\pi$. (Note that this is somewhat different from the parametrization chosen in \cite{Arean:2004mm}.) The embedding where the brane winds $m_1$ times around $\phi_1$ and $m_2$ times around $\phi_2$ can then be written as
\begin{equation}
 \begin{gathered}
  \T=\T(\tau) \, , \qquad \R=\R(\tau) \, , \qquad X_i=x^0_i \quad (i=1,2,3) \, , \\
  \Theta_1 = 2\tan^{-1}(c_1\zeta^{m_1}) \, , \qquad \Phi_1 = m_1\beta \, , \\ 
  \Theta_2 = 2\tan^{-1}(c_2\zeta^{m_2}) \, , \qquad \Phi_2 = m_2\beta \, , \\
  \Psi= \gamma \ .
 \end{gathered}
\end{equation}
Here, $x^0_i$ are arbitrary constants representing the location of the wrapped brane in the field theory directions --- these are unimportant due to translational invariance. Moreover, $c_1$ and $c_2$ are arbitrary positive constants. When the winding number $m_1$ is zero, $c_1$ sets the $\theta_1$-coordinate of the brane, and similarly for $c_2$. The somewhat peculiar form of $\Theta_1$ and $\Theta_2$ ensures that
\begin{equation}
 \left(c_1^{-1}\tan{\frac{\Theta_1}{2}}\right)^{m_2} = \left(c_2^{-1}\tan{\frac{\Theta_2}{2}}\right)^{m_1}
\end{equation}
holds, which is our version of (3.23) in \cite{Arean:2004mm}; to compare, set $m_1=1$, $m_2=m$.\footnote{Note that the brane could have some non-zero angular velocity (and associated angular momentum) in the $\phi$-directions (one linear combination of angular velocities in the $(\phi_1,\phi_2)$-directions is pure gauge, but there remains a physical velocity as well). However, it seems obvious that this will only increase the energy, and so we set it to zero here. We have explicitly checked this for $(m_1,m_2)=(1,0)$ and $(0,1)$.} We can visualize this embedding with the following diagram:
\beq
\begin{array}{rccccccccccl}
 &t & x & y & z & r & \theta_1 &\phi_1  &\theta_2 &\phi_2 & \psi & \nonumber \\
N_c\text{ color D3 (background)}: &\otimes & \otimes &\otimes &\otimes &\cdot &\cdot & \cdot &\cdot &\cdot &\cdot &   \\
\text{Baryonic probe D3, }m_1\neq0,\ m_2=0: &\otimes &\cdot & \cdot &\cdot &\cdot & \otimes & \otimes &\cdot &\cdot &\otimes & \\
\text{Baryonic probe D3, }m_1=0,\ m_2\neq0: &\otimes &\cdot & \cdot &\cdot &\cdot &\cdot & \cdot &\otimes &\otimes &\otimes & \\
\text{Baryonic probe D3, }m_1\neq0,\ m_2\neq0: &\otimes &\cdot & \cdot &\cdot &\cdot &\oslash &\oslash &\oslash &\oslash &\otimes & \\
\end{array}
\eeq
According to a straightforward generalization of the arguments from \cite{Arean:2004mm}, this embedding should be dual to a field theory operator of the schematic form
\begin{equation}\label{eq:baryonOp}
 \left(A^{|m_1|} B^{|m_2|}\right)^{N_c} \ ,
\end{equation}
with conformal dimension $3(|m_1|+|m_2|)N_c/4$ and baryon number $(|m_1|-|m_2|)N_c$.

If we take $\dot \R=0$, the brane is not moving at all in our embedding, and the effective potential for these baryonic D3-branes is essentially given by minus their on-shell Lagrangian. We again start from the D3-brane action \eqref{eq:coloraction}. To evaluate it on the above baryonic embedding, we go to static gauge where $\T=\tau$ and integrate the resulting action over the internal directions. The pullback of $C_4$ in (\ref{eq:C4}) becomes
\begin{equation}
\begin{split}
 P[C_4] = -\frac{\sqrt{2}}{9}\Bigg\{& \left( \frac{c_1^2\, m_1^2\, \zeta^{2m_1-1}}{\left(1+c_1^2\, \zeta^{2m_1}\right)^2} - \frac{c_2^2\, m_2^2\, \zeta^{2m_2-1}}{\left(1+c_2^2\, \zeta^{2m_2}\right)^2} \right)\Phi \\
 + &\left( \frac{c_1^2\, m_1^2\, \zeta^{2m_1-1}}{\left(1+c_1^2\, \zeta^{2m_1}\right)^2} + \frac{c_2^2\, m_2^2\, \zeta^{2m_2-1}}{\left(1+c_2^2\, \zeta^{2m_2}\right)^2} \right) \left(\Phi_M - \Phi_R\right) \Bigg\} d\tau \wedge d\zeta \wedge d\beta \wedge d\gamma \ ,
\end{split}
\end{equation}
while the DBI term evaluates to
\begin{equation}
 \sqrt{-\det(P[g_{\mu\nu}])} = \frac{2}{9}e^{-\frac{w}{2}-\eta+\frac{2\chi}{3}}\sqrt{g} \left( \frac{c_1^2\, m_1^2\, \zeta^{2m_1-1}}{\left(1+c_1^2\, \zeta^{2m_1}\right)^2} e^{\lambda} + \frac{c_2^2\, m_2^2\, \zeta^{2m_2-1}}{\left(1+c_2^2\, \zeta^{2m_2}\right)^2} e^{-\lambda} \right) \ .
\end{equation}
Integrating over the internal directions is fairly simple; the integrals over $\beta$ and $\gamma$ are trivial and give a factor of $8\pi^2$, while the integral over $\zeta$ takes the form
\begin{equation}
 \int_0^\infty d\zeta\, \frac{c^2\, m^2\, \zeta^{2m-1}}{\left(1+c^2\, \zeta^{2m}\right)^2} = \frac{|m|}{2} \ .
\end{equation}
The full action then becomes
\begin{equation}
\begin{split}
 S_{bD3} = - T_3\, 8\pi^2 \int d\tau \bigg\{& \frac{1}{9} e^{-\frac{w}{2}-\eta+\frac{2\chi}{3}}\sqrt{g} \left( |m_1| e^{\lambda} + |m_2| e^{-\lambda} \right) \\
 &+ \epsilon_3 \left( |m_1| - |m_2| \right)\frac{\Phi}{9\sqrt{2}} + \epsilon_3 \left( |m_1| + |m_2| \right)\frac{\Phi_M - \Phi_R}{9\sqrt{2}} \bigg\} \equiv \int d\tau L_{bD3} \ ,
\end{split}
\end{equation}
where we have defined the effective Lagrangian $L_{bD3}$. Since all kinetic terms are set to zero, the effective potential is just minus the Lagrangian,
\begin{equation}
\begin{split}
 V_{bD3} \equiv -\frac{1}{T_3 V_{S^3}}L_{bD3} = & \frac{1}{18} e^{-\frac{w}{2}-\eta+\frac{2\chi}{3}}\sqrt{g} \left( |m_1| e^{\lambda} + |m_2| e^{-\lambda} \right) \\
 &+ \epsilon_3 \left( |m_1| - |m_2| \right)\frac{\Phi}{18\sqrt{2}} + \epsilon_3 \left( |m_1| + |m_2| \right)\frac{\Phi_M - \Phi_R}{18\sqrt{2}} \ , \label{eq:bD3effPotGeneral}
\end{split}
\end{equation}
where $V_{S^3}=16\pi^2$. Note that $|m_1| - |m_2|$ sets the coupling to the baryonic gauge field $\Phi$, while $|m_1| + |m_2|$ sets the coupling to the $R$-charge gauge field $\Phi_R$, in agreement with \eqref{eq:baryonOp}. Moreover, when $\lambda=0$, $|m_1| + |m_2|$ is proportional to the mass as well as the $R$-charge. This is a consequence of the field theory being superconformal in which case the $R$-charge and the conformal dimension of the corresponding operator are related.

Expanding (\ref{eq:bD3effPotGeneral}) for large radius we get
\begin{equation}
 V_{bD3} = \frac{|m_1| + |m_2|}{18}\, r + \epsilon_3 \left( |m_1| - |m_2| \right)\frac{\Phi_{0,0}}{18\sqrt{2}} - \epsilon_3 \left( |m_1| + |m_2| \right)\frac{\Phi_{R\,0,0}}{18\sqrt{2}} + \mathcal{O}(r^{-1}\log{r}) \ ,
\end{equation}
which shows that the potential always grows linearly in $r$ for any non-trivial winding, and thus the global minimum of the potential cannot be at the asymptotic boundary.

We see from (\ref{eq:bD3effPotGeneral}) that while we can indeed get operators with arbitrarily large baryon charge, for example by taking $m_2$ large, these operators will also be very massive. Taking, {\emph{e.g.}}, $m_2$ large and $m_1=0$ just multiplies the potential with an overall $|m_2|$, which cannot change the sign of the potential to create a minimum away from the horizon. Since we should take $\epsilon_3=+1$ and $\Phi>0$ in the bulk (while $\Phi_M-\Phi_R <0$ mostly), one possibility of getting such a minimum could be if we took $m_2\neq 0$ and $m_1=0$ in a background with $\lambda$ large in the bulk. The large $\lambda$ in the bulk makes the $S^2$ the brane wraps small, and thus the mass remains small while the baryon charge is unaffected by $\lambda$. In this case it might be possible that the potential changes sign. Unfortunately we have not been able to construct backgrounds with $\lambda$ large enough to study this possibility.

\begin{figure}[h!]
\begin{center}
\includegraphics[scale=0.56,valign=t]{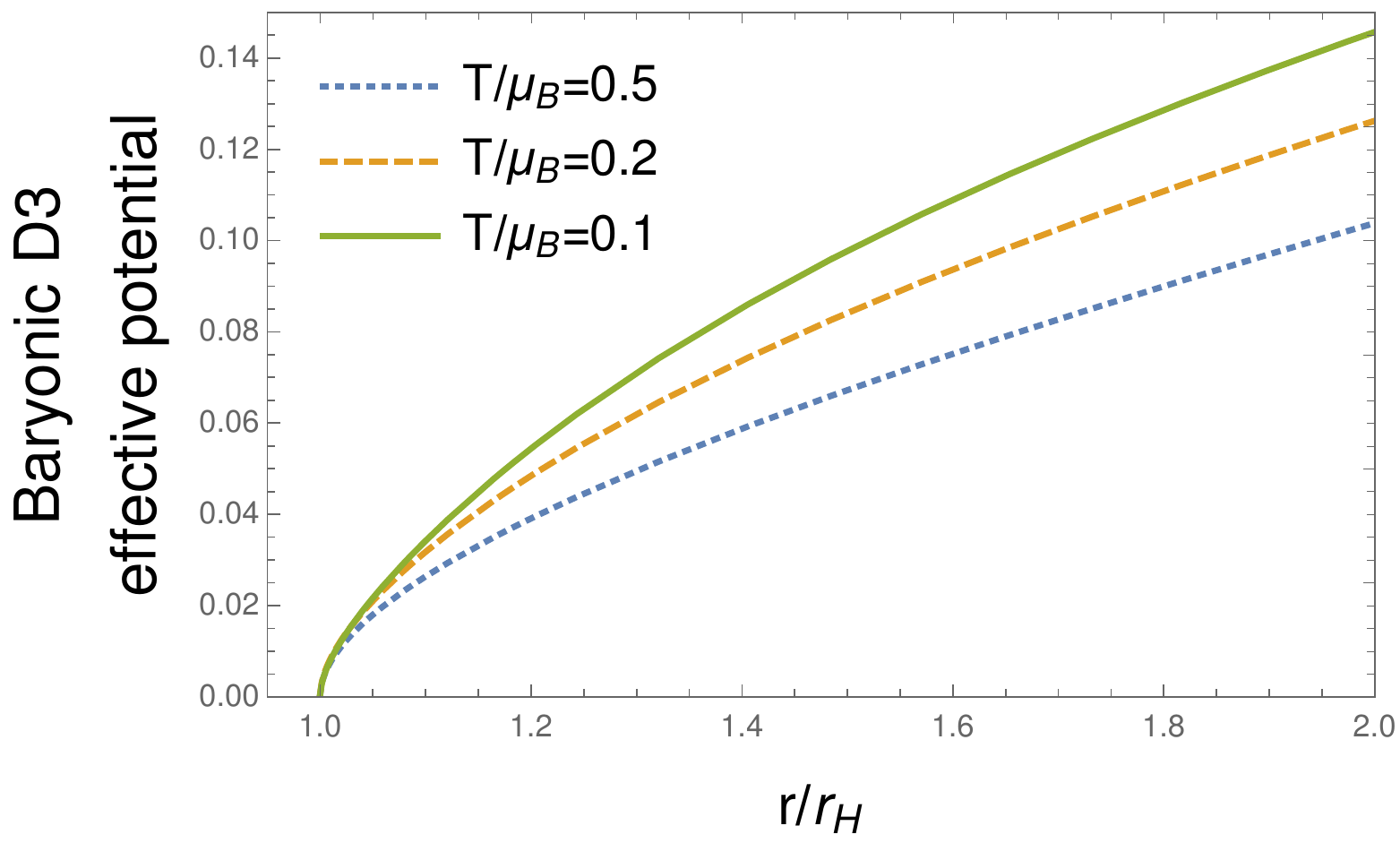}
\includegraphics[scale=0.52,valign=t]{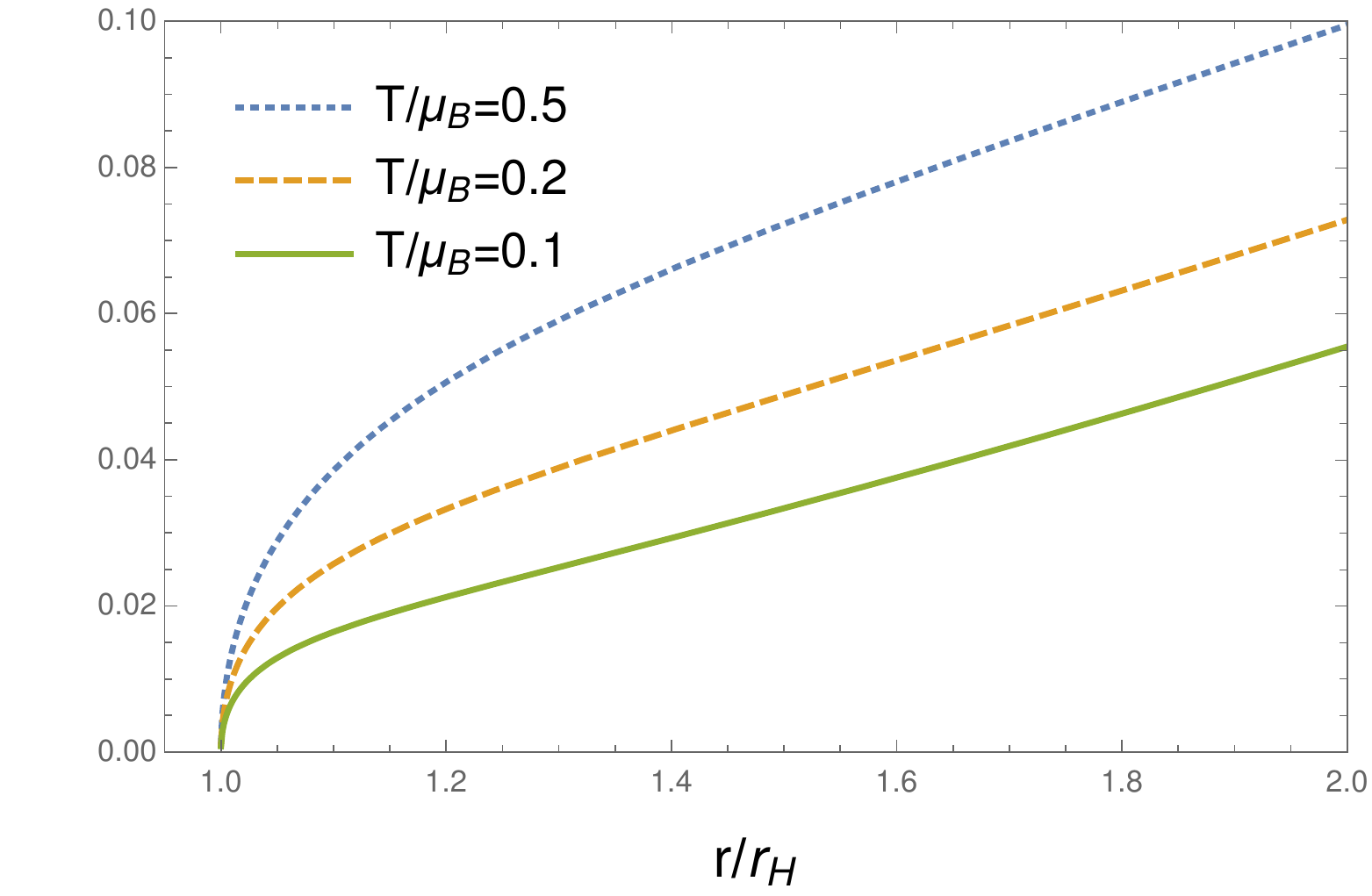}
\caption{The effective potential for a probe baryonic D3-brane wrapped around the $S^3$ parametrized by $\{\theta_1,\phi_1,\psi\}$ (left) and $\{\theta_2,\phi_2,\psi\}$ (right). The curves are all at $T/M=0.5$, with varying $T/\mu_B$.}\label{fig:baryonicD3effPot}
\end{center}
\end{figure}

We have depicted the effective potential (\ref{eq:bD3effPotGeneral}) as a function of the radial coordinate in Fig.~\ref{fig:baryonicD3effPot} for both embeddings. In this figure, we keep the mass of the fermions and the temperature fixed and vary the chemical potential. We find that at high densities the potential is flattening for the embedding wound around the  $\{\theta_2,\phi_2,\psi\}$ directions. The results are qualitatively similar for any $T/M$. 

\begin{figure}[h!]
\begin{center}
\includegraphics[scale=0.62,valign=t]{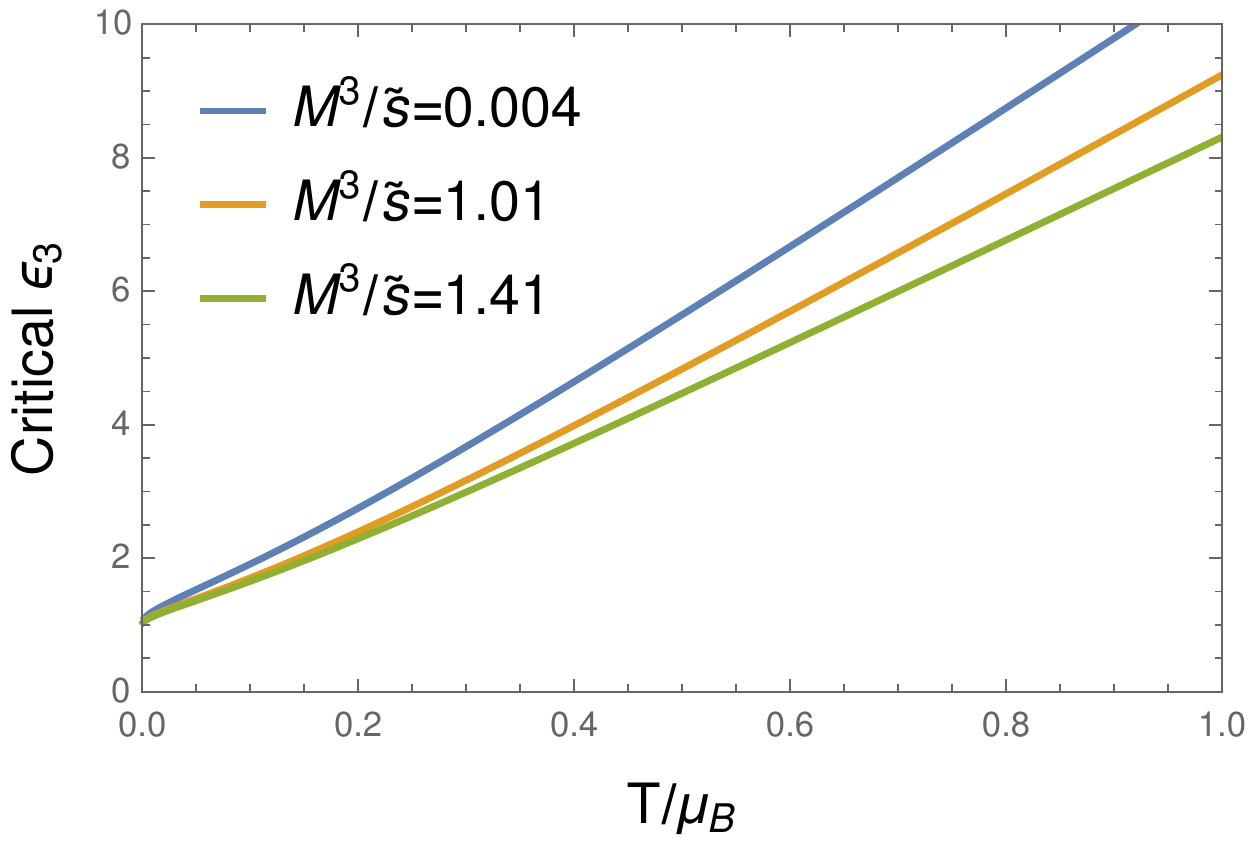}
\includegraphics[scale=0.61,valign=t]{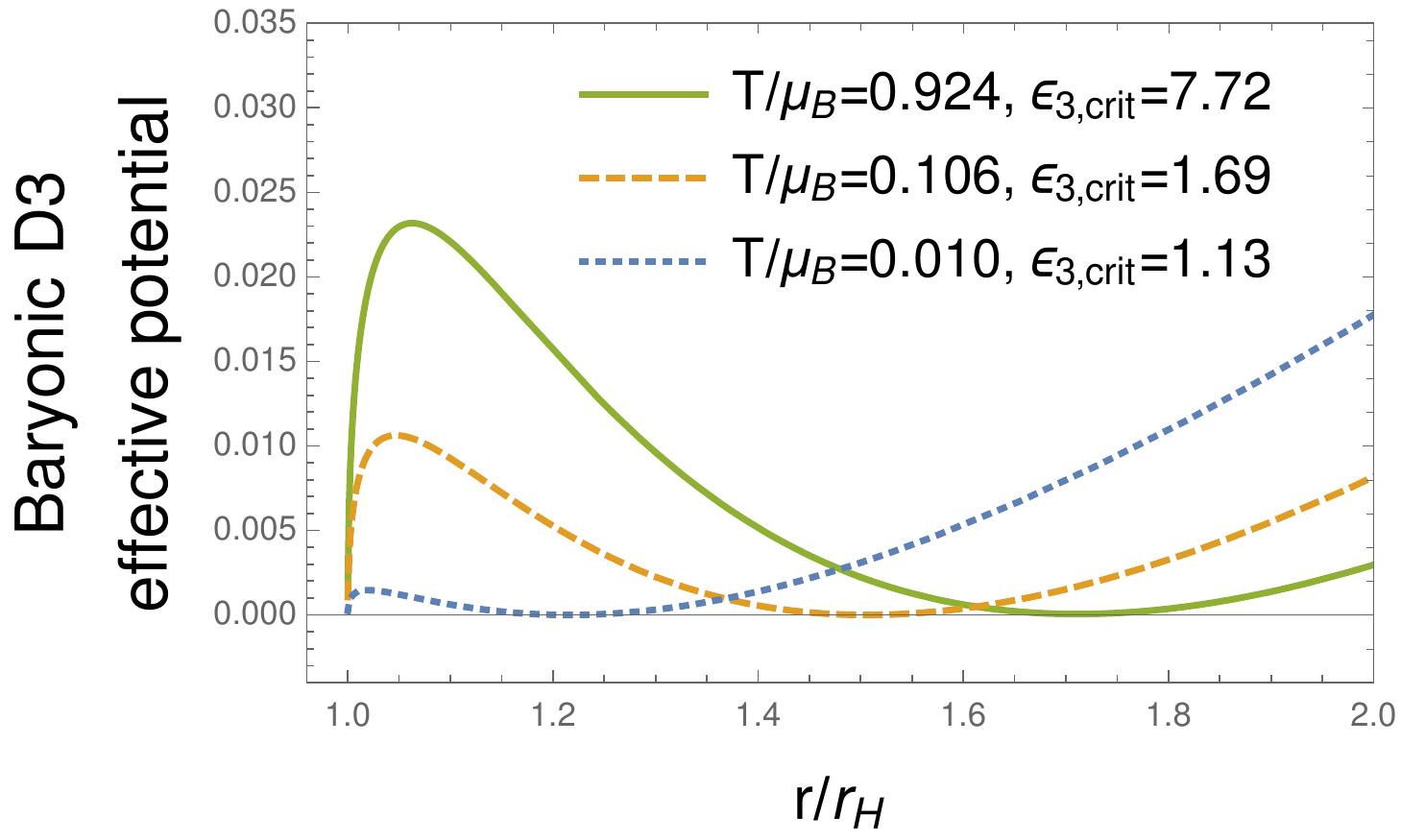}
\caption{Left: The critical value of $\epsilon_3$ where baryon condensation could occur for the $(m_1,m_2)=(0,1)$ wrapping, as a function of $T/\mu_B$, for increasing values of $M^3$ normalized by $\tilde s\equiv G_5\, s$, where $s$ is the entropy density. Right: The baryonic potential along the $M^3/\tilde s=1.41$ curve on the left.}\label{fig:criticalEpsilon}
\end{center}
\end{figure}

While a minimum outside the horizon does not form for the baryons, it may be worth noting that this would occur if we allow ourselves to take the D3-brane charge $\epsilon_3$ slightly larger. This of course takes us away from the strict top-down framework, but is nonetheless useful to study. For zero deformation, it was demonstrated in \cite{Herzog:2009gd} that if $\epsilon_3>1$ baryon condensation would occur at low enough $T/\mu_B$. In the left panel of Fig.~\ref{fig:criticalEpsilon}, we show the smallest value of $\epsilon_3$ for which condensation occurs as a function of $T/\mu_B$, for three different values of $M^3/\tilde s$, where\footnote{In units where the aAdS radius is $L=1$.} $\tilde s= G_5 s$ (the results of \cite{Herzog:2009gd} correspond to $M^3/\tilde s=0$). It appears that all curves hit $\epsilon_3=1$ when $T$ goes to zero and grow monotonically with $T$. In the right panel of the same figure, we show the potential along the bottom curve of the left plot. As we approach zero temperature and the critical $\epsilon_3$ approaches 1, the minimum of the potential approaches the horizon while asymptotically we still see linear growth. Together, this indicates that no condensation occurs throughout the accessible phase diagram, even in the zero temperature limit. Furthermore, the extremal $T=0$ geometry appears to be marginally stable: There is a minimum right by the horizon, which for any $\epsilon_3>1$ would dip below zero leading to an instability.

\section{Discussion} \label{sec:discussion}

We have considered a deformation of the Klebanov-Witten theory that breaks both conformal invariance and supersymmetry. Since the classical potential in the deformed theory is unbounded from below, the theory should be unstable at low temperature. This is confirmed by the analysis of the gravity dual, albeit the instability seems to be realized differently depending on the value of the baryon chemical potential. For small values of the baryon chemical potential it is manifested directly in the properties of the classical gravity solutions. In particular, stable solutions fall outside a region of the phase diagram limited by the points where the speed of sound in the field theory dual goes to zero and the expectation value of a scalar operator diverges. For larger values of the baryon chemical potential the gravity solutions are classically stable, but the effective potential of probe color branes has a global minimum outside the horizon at low temperatures, thus there is a brane nucleation instability. Interestingly, the on-set of this nucleation instability is independent of the conformal symmetry breaking scale. We have thus established that in the extremal limit all the charged solutions are either unstable or metastable. 

The results are in principle in accord with the weak gravity conjecture, but the realization of the instability seems to deviate from the usual argument. The black brane has baryon charge (while the $R$-charge is vanishing) and one might have expected that the instability of extremal branes would be related to baryonic branes, but instead color branes are the ones showing the nucleation instability. It is conceivable that the nucleation instability we observe is unrelated to the weak gravity conjecture. A possible way to check the conjecture more directly would be to modify the potential in the field theory dual in such a way that it remains bounded from below. In the setup we have studied this could be partially accomplished by introducing a double-trace deformation that is the square of the $\Delta=2$ operator
\be
 \sim \left({\rm Tr}\,\left( |a|^2-|b|^2\right)\right)^2.
\ee
This can be realized in the gravity dual by imposing mixed boundary conditions for the dual field $\lambda$ \cite{Witten:2001ua}. However, it is not enough because the potential remains flat along the $|a|^2=|b|^2$ directions, so chemical potentials will generically make the effective potential unbounded from below. Then, the nucleation instability of color branes seems unavoidable as long as one remains close to the classical supergravity limit and may render extremal black holes unstable even without invoking the weak gravity conjecture. Most theories with known gravity duals are supersymmetric or they are deformations of supersymmetric theories with moduli spaces, so this situation is probably quite general. If it was possible to remove the color brane nucleation instability, in principle an instability related to the nucleation of baryonic branes may emerge once quantum corrections are taken into account on the gravity side ($1/N_c$ corrections on the field theory side). On the other hand, it could as well be that quantum corrections do not trigger a new instability but that the nucleation instability of color branes is the mechanism by which many theories with holographic duals avoid the issues related to extremal black holes in quantum gravity.

One of our original motivations to study the Klebanov-Witten model is that it might be used as a toy model for QCD with non-zero baryon charge, realized without having to introduce flavor branes with their associated technical complications. In addition, by considering larger consistent truncations \cite{Cassani:2010na}, the Klebanov-Witten theory can be deformed to Klebanov-Strassler \cite{Klebanov:2000hb}, that shows confinement and is thus even more similar to QCD. In this context the results related to D5-branes carrying color D3-brane flux are particularly interesting. We have found that the effective potential has a global minimum at a finite distance outside the horizon, so in principle metastable configurations with D5-branes localized at the minimum can exist and will correspond to a partial Higgsing of the gauge group. With obvious differences, this is akin to color superconductivity in QCD and it would be very interesting to explore further. The Higgsed phase could perhaps be used to model matter in the deep cores of the most compact objects in the universe. Therefore, results in this direction goes beyond academic curiosity and will help us understand the exotic phases of matter pressed in immense pressures, ultimately probed in astronomical laboratories of merging neutron stars. To our knowledge the only other example with similar features is $\cN=4$ SYM compactified on a sphere  \cite{Henriksson:2019zph}, which is less interesting as there is no baryon charge and metastable configurations go away in the flat space limit. 

On a related direction, a deeper understanding of the instabilities on the gravity side may also led to new interesting realizations of ordered phases. Toy versions of Chern-Simons driven instabilities have already led to increasing understanding of breaking of continuous symmetries in various dimensions \cite{Nakamura:2009tf,Bergman:2011rf}. It would likewise be interesting to understand the underlying mechanism behind the Chern-Simons terms that lead to color superconducting ground states. In this vein, it would be possible to demonstrate the precursor mechanism to transition to such phases in some realistic holographic models for quantum chromodynamics, {\emph{e.g.}}, in V-QCD \cite{Jarvinen:2011qe}.


\addcontentsline{toc}{section}{Acknowledgments}
\paragraph{Acknowledgments}

\noindent

We would like to thank Alfonso V. Ramallo, Javier Tarr\'io, and Aleksi Vuorinen for discussions, and Javier Subils for advice on how to improve our numerical algorithms. O.~H. is supported by the Academy of Finland grants no. 1297472 and 1322307, as well as by a grant from the Ruth and Nils-Erik Stenb\"ack foundation. C.~H. is partially supported by the Spanish grant PGC2018-096894-B-100 and by the Principado de Asturias through the grant FC-GRUPIN-IDI/2018/000174. N.~J. is supported in part by the Academy of Finland grant no. 1322307.

\appendix


\section{SUGRA formulas}\label{app:sugra}

In this section we will specify the supergravity (SUGRA) background geometry which provides us with the dual of the gauge theory discussed in the introductory part of this paper. We  will start with the full 10D SUGRA Lagrangian and lay out the Ansatz for the metric and for the various fluxes. However, rather than directly solving for the 10D fields, we will simplify our task by reducing the problem to a lower 5-dimensional effective field theory that is better fit for numerical analysis. The pay-off is that there are more fields to be solved for, {\emph{e.g.}}, the non-trivial radially dependent 10D metric will imply various scalar fields in the 5D case.

\subsection{10D theory}

The conifold gauge theory lives on the worldvolume of a stack of D3-branes placed at the tip of the conifold $T^{1,1}$. By taking the usual near-horizon limit, it can be shown to be dual to type IIB string theory on $AdS_5\times T^{1,1}$. In the appropriate limit, this reduces to IIB supergravity, whose action is
\begin{equation}
 S_{IIB} = \frac{1}{2\kappa_{10}^2} \int \left[ R - \frac{1}{2}(d\phi)^2 - \frac{1}{2}e^{-\phi}H^2 - \frac{1}{2}e^{2\phi}(F_1)^2 - \frac{1}{2}e^{\phi}(F_3)^2 - \frac{1}{4}(F_5)^2 \right] * 1 + CS
\end{equation}
Following \cite{Herzog:2009gd} we will study truncations of this supergravity action where all fields except the metric and the RR 5-form are set to zero. Our solution will be written in terms of three vector fields: $A$, associated to the baryon charge, $A_R$, associated to the $R$-charge, and a massive vector field $A_M$. The corresponding field strengths are $F = d A$, $F_R = d A_R$, and $F_M = d A_M$. We define the following forms on $T^{1,1}$:
\begin{equation}
\begin{split}
\omega_2 &\equiv {1\over 2} \left(\sin \theta_1 d\theta_1 \wedge d\phi_1
- \sin \theta_2 d\theta_2 \wedge d\phi_2 \right)  \\
\omega_3 &\equiv g_5 \wedge \omega_2  \\
g_5 &\equiv d \psi + \cos \theta_1 d\phi_1 + \cos \theta_2 d\phi_2 \\
g_5^A &\equiv g_5 + {3 \over \sqrt{2}} A_R \ .
\end{split}
\end{equation}
The Ansatz for the 10D metric is
\bea
 ds_{10}^2 & = & e^{-5 \chi/3} ds_5^2 + e^{\chi}
\bigg[{e^{\eta + \lambda} \over 6} (d \theta_1^2 + \sin^2 \theta_1 d\phi_1^2) \nonumber\\
& & + {e^{\eta-\lambda} \over 6} (d \theta_2^2 + \sin^2 \theta_2 d\phi_2^2)+ {e^{-4 \eta} \over 9} \left(g_5 + {3 \over \sqrt{2}} A_R\right)^2 \bigg] \ ,\label{eq:10Dmetric}
\eea
where $ds_5^2$ is a 5D line element (with AdS$_5$ asymptotics). Note that if the scalar fields and $A_R$ are set to zero the terms in square brackets above reduce to the canonical metric on $T^{1,1}$ \cite{Herzog:2009gd}. It is convenient to define the 3-forms
\begin{align}
 \mathcal{Q}_B^{(3)} &\equiv e^{2\eta - \frac{4\chi}{3}} \left[ \cosh(2\lambda)\star_5 F - \sinh(2\lambda)\star_5 \left( F_M - F_R \right) \right] \\
 \mathcal{Q}_R^{(3)} &\equiv e^{2\eta - \frac{4\chi}{3}} \left[ \cosh(2\lambda)\star_5 \left( F_R - F_M \right) + \sinh(2\lambda)\star_5 F \right] + \frac{1}{2}e^{-4\eta + \frac{8\chi}{3}} \star_5 F_R \\
 \mathcal{Q}_M^{(3)} &\equiv e^{2\eta - \frac{4\chi}{3}} \left[ \cosh(2\lambda)\star_5 \left( F_M - F_R \right) - \sinh(2\lambda)\star_5 F \right] \ ,
\end{align}
where $\star_5$ denotes the Hodge dual with respect $ds_5$. The first two of these are related to the conserved baryon and $R$-charge densities of the dual field theory, respectively. With the help of these 3-forms, the self-dual 5-form can be written as
\begin{equation}\label{eq:5form}
\begin{split}
F_5 =& {1 \over g_s} \left( {\cal F} + *{\cal F} \right)  \\
{\cal F} =& -{2 \over 27} \omega_2 \wedge \omega_2 \wedge g_5 - {1 \over 9 \sqrt{2}} F \wedge \omega_2 \wedge g_5^A  \\
&+ {1 \over 18 \sqrt{2}} \left( F_M - F_R \right) \wedge d g_5 \wedge g_5^A - {1 \over 18 \sqrt{2}} \left( A_M - A_R \right) \wedge dg_5 \wedge dg_5   \\
*{\cal F} =& -4 e^{-{20 \over 3}\chi} \mathrm{vol}_M - {1 \over 3 \sqrt{2}} \mathcal{Q}_B^{(3)} \wedge \omega_2 - {1 \over 6 \sqrt{2}} \mathcal{Q}_M^{(3)} \wedge dg_5 \\
&- {2 \sqrt{2} \over 3} e^{-4 \chi - 4 \eta} (\star_5 A_M) \wedge g_5^A \ .
\end{split}
\end{equation}

For probe brane computations, we need to find an explicit expression for the RR four-form potential $C_4$ satisfying $F_5=-\frac{2}{27} \omega _2\wedge \omega _2\wedge g_5 + dC_4$. We arrive at
\begin{equation}
\begin{split}
 C_4 =& -\frac{1}{9\sqrt{2}}A \wedge \omega_2 \wedge g_5^A + \frac{1}{18\sqrt{2}}\left( A_M - A_R \right) \wedge d g_5 \wedge g_5^A \\
 &+ \frac{1}{6\sqrt{2}}\mathcal{Q_M} \wedge g_5 + \frac{1}{3\sqrt{2}}\mathcal{Q_B} \wedge h + a_4(r) dt\wedge dx\wedge dy\wedge dz \label{eq:C4} \ .
\end{split}
\end{equation}
Here we have additionally defined the 1-form $h\equiv\cos{\theta_1}d\phi_1 - \cos{\theta_2}d\phi_2$. The function $a_4$ is determined by the equation
\begin{equation}
 a_4'(r) = 2r^3e^{-\frac{w}{2}-\frac{20\chi}{3}}\left( 2 - \frac{2e^{w-4\eta+\frac{8\chi}{3}}\Phi_R \Phi_M}{g} \right) \ ,
\end{equation}
together with the condition that it goes to zero at the horizon.

\begin{table}[!ht]\label{table:Cassani}
\begin{tabularx}{1.\textwidth}{lYcYcYYYc}  
\toprule
$\mathcal{N}=2$ multiplet && field fluctuations  && $m^2$ & $\phantom{s}$ & $\Delta$ & $\phantom{s}$ & dual operators  \\
\midrule
gravity && $\begin{array}{c} 3A_R-2A_M \\ g_{\mu\nu} \end{array}$ && $\begin{array}{c} 0 \\ 0 \end{array}$ && $\begin{array}{c} 3\\ 4 \end{array}$ &&  ${\rm Tr}(W_{1\alpha}\overline W_{\!1 \dot\alpha} + W_{2\alpha}\overline W_{\!2 \dot\alpha} )+\ldots $ \\ 
\midrule
Betti vector && $\begin{array}{c} \lambda \\ A \end{array}$ && $\begin{array}{c} -4 \\ 0\end{array}$ && $\begin{array}{c} 2\\ 3 \end{array}$ && ${\rm Tr}\,A e^{V_2} \overline A e^{-V_1} -{\rm Tr}\, B e^{V_1}\overline B e^{-V_2}$ \\ 
\midrule
massive vector && $\begin{array}{c} \eta \\ A_M \\ \chi \end{array}$ && $\begin{array}{c} 12\\ 24 \\ 32  \end{array}$  && $\begin{array}{c}  6\\   7 \\  8 \end{array}$ &&  ${\rm Tr}(W_{\!1}^2\overline W{}_{\!1}^2 + W_{\!2}^2\overline W{}_{\!2}^2) + \ldots $ \\
\bottomrule
\end{tabularx}
\caption{Mass eigenstates of the type IIB supergravity truncation introduced here on the supersymmetric AdS$_5\times T^{1,1}$ background, and their dual superfield operators, adapted from \cite{Cassani:2010na}.}
\end{table}

Plugging all this into the 10D action and integrating over the compact conifold dimensions, one obtains the following 5D action:
\begin{equation}\label{eq:HerzogLag}
 S_{5D} = \frac{1}{16\pi G_5}\int d^5 x \sqrt{-g} \mathcal{L}_{5D} + S_{CS} \, ,
\end{equation}
where
\bea
 \mathcal{L}_{5D}  & = & R - \frac{10}{3}(\partial_{\mu}\chi)^2 - 5(\partial_{\mu}\eta)^2 - (\partial_{\mu}\lambda)^2 - V \nonumber\\
 & & -\frac{1}{4}e^{2\eta-\frac{4}{3}\chi}\left[ \cosh(2\lambda) \left( (F_{\mu\nu})^2 + (F^M_{\mu\nu}-F^R_{\mu\nu})^2 \right) - 2\sinh(2\lambda)(F^M_{\mu\nu}-F^R_{\mu\nu})F^{\mu\nu} \right] \nonumber\\
 & & -\frac{1}{8}e^{-4\eta+\frac{8}{3}\chi}(F^R_{\mu\nu})^2 - 4e^{-4\eta-4\chi}(A^M_{\mu})^2 \, ,
\eea
and the potential is
\begin{equation}\label{eq:scalarPotential}
 V = 8e^{-\frac{20}{3}\chi} + 4e^{-\frac{8}{3}\chi} (e^{-6\eta}\cosh(2\lambda)-6e^{-\eta}\cosh(\lambda)) \, ;
\end{equation}
we have set the radius of curvature $L=1$. The Chern-Simons term in five dimensions reads
\begin{equation}
 S_{CS} = \frac{1}{2\sqrt{2}} \int (A_M - A_R) \wedge F_M \wedge F_R - \frac{1}{2\sqrt{2}} \int A \wedge F \wedge F_R \, .
\end{equation}
For completeness, we have included the Table~\ref{table:Cassani}, which summarizes the operator duals and their dimensions of the corresponding SUGRA fields.


\section{Boundary analysis and holographic renormalization}\label{app:holoren}

In this section we will carefully discuss the necessary holographic renormalization which we have performed in order to extract the thermodynamics for the background. Though the methods we use are standard \cite{Skenderis:2002wp}, we encounter several subtleties with logarithms, and the associated renormalization scheme. Therefore, we prefer to essentially follow the conventions set in \cite{Hoyos:2016cob,Ecker:2017fyh}, where also a much more expanded discussion can be found.

\subsection{Boundary expansion}

For large $r$, near the aAdS boundary, we can solve the equations of motion order by order. To do this, we make an Ansatz for all the fields of the form
\begin{equation}
 F(r) = \sum_{i\geq0}\sum_{j\geq0}F_{i,j}r^{-i} \log(r/L)^j \ ,
\end{equation}
We have assumed that the sources of the irrelevant operators dual to $\Phi_M$, $\eta$, and $\chi$ are set to zero. The result is as follows:
\begin{equation}
\begin{split}\label{eq:UVexpansion}
 g &= \frac{r^2}{L^2} + \frac{L^2}{r^2}\left(g_{2,0} + \frac{\log(r/L)}{3}(4\lambda_{2,1}\lambda_{2,0}-\lambda_{2,1}^2) + \frac{\log(r/L)^2}{3}2\lambda_{2,1}^2 \right) + \mathcal{O}(r^{-4}) \\
 \Phi &= \Phi_{0,0} + \frac{L^2\Phi_{2,0}}{r^2} + \mathcal{O}(r^{-4}) \ , \qquad \Phi_R = \Phi_{R\,0,0} + \frac{L^2\Phi_{R\,2,0}}{r^2} + \mathcal{O}(r^{-4}) \\
 \Phi_M &= \mathcal{O}(r^{-4}) \ , \qquad \eta = \mathcal{O}(r^{-4}) \ , \qquad \chi = \mathcal{O}(r^{-4}) \ , \qquad w = \mathcal{O}(r^{-4}) \\
 \lambda &= \frac{L^2\lambda_{2,0}}{r^2} + \frac{L^2\lambda_{2,1}\log(r/L)}{r^2} + \mathcal{O}(r^{-4}) \ .
\end{split}
\end{equation}
We see that rescaling of the argument of the log $L\rightarrow sL$ causes, in particular, a shift $\lambda_{2,0}\rightarrow\lambda_{2,0}-\lambda_{2,1}\log(s)$. This corresponds to a change of scheme, in the following we will fix $s=1$. Note that there appears three more independent constants at subleading orders, related to the expectation values of the operators dual to $\Phi_M$, $\eta$, and $\chi$ --- these will not be important in the rest of our analysis.

\subsection{On-shell action and counterterms}

We want to show that, with the help of the equations of motion, the action can be written as a total derivative, greatly simplifying its evaluation. First, by considering the trace of Einstein equations, it is easy to show that the on-shell action can be written as
\begin{equation}
 S_{OS} = \frac{1}{16\pi G_5}\int d^5x\sqrt{-g}\frac{2}{3}\left[ V + F_{kin} \right] \ ,
\end{equation}
where $V$ is the scalar potential (\ref{eq:scalarPotential}) and $F_{kin}$ contains the kinetic terms for the vector fields:
\begin{equation}
 F_{kin} \equiv - \frac{1}{4}e^{2\eta-\frac{4}{3}\chi}\left[ \cosh(2\lambda) \left( (F_{\mu\nu})^2 + (\tilde F_{\mu\nu}^R)^2 \right) - 2\sinh(2\lambda)F_{\mu\nu} \tilde F_R^{\mu\nu} \right]
 -\frac{1}{8}e^{-4\eta+\frac{8}{3}\chi}(F^R_{\mu\nu})^2 \ .
\end{equation}
With further use of the Einstein equations one also finds
\begin{equation}
 V + F_{kin} = \frac{3}{r}\left(-g'(r) + \frac{g(r)}{2}w'(r) - \frac{2g(r)}{r} \right) \ , 
\end{equation}
which lets us write
\begin{equation}
 S_{OS} = \frac{1}{16\pi G_5}\int d^5x\sqrt{-g}\frac{2}{r}\left[ -g'(r) + \frac{g(r)}{2}w'(r) - \frac{2g(r)}{r} \right] \ .
\end{equation}
By evaluating $\sqrt{-g}$ it is easy to combine these terms into a total derivative and arrive at the final expression
\begin{align}
 S_{OS} &= \frac{1}{16\pi G_5}\int d^5 x \, \partial_r \left[ -2\sqrt{-g}\frac{g(r)}{r} \right] \\
 &= -\frac{1}{16\pi G_5}\int d^4 x \left[2\sqrt{-g}\, \frac{g(r)}{r} \right]_{r=r_H}^{r=r_\Lambda} \ ,
\end{align}
where we performed the radial integral from the horizon $r=r_H$ to a cutoff near the boundary $r_\Lambda$.

As always, one needs to add to this a Gibbons-Hawking boundary term
\begin{equation}
 S_{GH} = \frac{1}{8\pi G_5}\int d^4x \sqrt{-\gamma}K \ ,
\end{equation}
where $\gamma$ is the determinant of the induced metric and $K$ is the extrinsic curvature. Furthermore, one needs a boundary ``cosmological constant'' term
\begin{equation}
 S_\Lambda = -\frac{1}{16\pi G_5}\int d^4x \sqrt{-\gamma}\Lambda
\end{equation}
to cancel out the volume divergence. Lastly, we need counterterms involving the scalar field $\lambda$:
\begin{equation}
 S_\lambda = \frac{1}{16\pi G_5}\int d^4x \sqrt{-\gamma}\left[ \left(c_1 + \frac{c_2}{\log(r/L)}  +  \frac{c_3}{\log(r/L)^2}\right) \lambda^2 \right] \ .
\end{equation}
The full on-shell action then takes the form
\begin{equation}
 S = S_{OS} + S_{GH} + S_\Lambda + S_\lambda \ .
\end{equation}
We fix the constants $c_i$ in $S_\lambda$ by requiring that the on-shell action is finite, including the counterterms, as $r_\Lambda\rightarrow\infty$. This results in the requirement
\begin{align}
 c_1 = -2 \ , \qquad c_2 = 1 \ ,
\end{align}
leaving us with one unfixed coefficient. By varying the action with respect to the scalar field and inserting the boundary expansion, we can now determine the boundary expectation value
\begin{equation}
 \langle\mathcal{O}_\lambda\rangle \equiv \frac{1}{16\pi G_5}\frac{\delta S}{\delta \lambda}\propto 2( \lambda_{2,0} + c_3 \lambda_{2,1} ) \ .
\end{equation}
Without $c_3$, we would recover the naive result depending only on $\lambda_{2,0}$. Inclusion of $c_3$ shifts the expectation value $\langle\mathcal{O}_\lambda\rangle$ by a constant proportional to $\lambda_{2,1}$. This is reminiscent of the shift caused by a change of scheme explained earlier.  In fact, the value of $c_3$ is scheme-dependent, so if we change to a scheme with $s\neq 1$, we have to  shift $c_3\to c_3 +\log(s)$, in such a way that the expectation value of the scalar and other physical observables are scheme-independent.

\subsection{Stress tensor and conserved currents}

With the counterterms fixed we can compute the stress tensor as usual in holography. We vary the action with respect to the boundary induced metric. From this variation, we get the standard result 
\begin{equation}
 T_{ij} \equiv \frac{1}{16\pi G_5}\frac{\delta S}{\delta \gamma_{ij}} = \frac{\sqrt{-\gamma}}{16\pi G_5}\left[ K_{ij} - \gamma_{ij}K - 3 \gamma_{ij} -\frac{\gamma_{ij}}{2}\left(-2 + \frac{1}{\log(r/L)}  +  \frac{c_3}{\log(r/L)^2}\right)\lambda^2\right] \ ,
\end{equation}
where $K_{ij}$ is the extrinsic curvature of the $r=\textrm{constant}$ hypersurface. Plugging in the near-boundary expansion, we get a result for the non-zero components of the field theory stress tensor in terms of the asymptotics of the gravity fields:
\begin{align}
 \langle T_{00}\rangle &= \frac{1}{16\pi G_5} \left( -3g_{2,0} + 2\lambda_{2,0}^2 - 2\lambda_{2,0}\lambda_{2,1} - c_3\lambda_{2,1}^2 \right) \\
 &= \frac{1}{16\pi G_5} \left( -3g_{2,0} + 2\lambda_{2,0}^2 - \frac{1}{2}\langle\mathcal{O}_\lambda\rangle\lambda_{2,1} - \lambda_{2,0}\lambda_{2,1} \right) \\
 \langle T_{11}\rangle = \langle T_{22}\rangle = \langle T_{33}\rangle &= \frac{1}{16\pi G_5}\left( -g_{2,0} + \frac{2}{3}\lambda_{2,0}^2 + \frac{2}{3}\lambda_{2,0}\lambda_{2,1} + \left(\frac{1}{3}+c_3\right)\lambda_{2,1}^2 \right) \\
 &= \frac{1}{16\pi G_5}\left( -g_{2,0} + \frac{2}{3}\lambda_{2,0}^2 + \frac{1}{2} \langle\mathcal{O}_\lambda\rangle\lambda_{2,1} - \frac{1}{3}\lambda_{2,0}\lambda_{2,1}+\frac{1}{3}\lambda_{2,1}^2 \right) \ .
\end{align}
The conformal anomaly can be computed from this as
\begin{equation}
 T_{ij} \eta^{ij} = \frac{1}{16\pi G_5}\left( 4\lambda_{2,0}\lambda_{2,1} + (1+4c_3)\lambda_{2,1}^2 \right) = \frac{1}{16\pi G_5}\left( 2\langle\mathcal{O}_\lambda\rangle\lambda_{2,1} + \lambda_{2,1}^2 \right) \ .
\end{equation}

We also note that by varying the on-shell action with respect to the two massless gauge fields, one obtains the quantities dual to the conserved baryon and $R$-charge currents in the field theory. The time components of these currents agree with the quantities $\mathcal{Q}_B$ and $\mathcal{Q}_R$ defined in \eqref{eq:QBQR}, which are conserved under radial translations in the bulk.

\bibliographystyle{JHEP}
\bibliography{refs}

\end{document}